\documentclass[aps,floatfix,nofootinbib]{revtex4}

\usepackage[T1]{fontenc}
\usepackage{color}
\usepackage{longtable} 
\usepackage[latin9]{inputenc}
\usepackage{hyperref}

\usepackage{array}      

\usepackage{textcomp}
\usepackage{amstext}
\usepackage{amssymb}
\usepackage{graphicx}
\usepackage{setspace}

\usepackage{graphicx,bm,psfrag, amsmath}

\newcommand{\beq}{\begin{equation}}
\newcommand{\eeq}{\end{equation}}
\newcommand{\beqa}{\begin{eqnarray}}
\newcommand{\eeqa}{\end{eqnarray}}
\newcommand{\ba}{\begin{array}}
\newcommand{\ea}{\end{array}}

\newcolumntype{L}[1]{>{\raggedright\let\newline\\\arraybackslash\hspace{0pt}}m{#1}} 
\newcolumntype{C}[1]{>{\centering\let\newline\\\arraybackslash\hspace{0pt}}m{#1}}   
\newcolumntype{R}[1]{>{\raggedleft\let\newline\\\arraybackslash\hspace{0pt}}m{#1}}  

\begin{document}
	\setstretch{1.2}

	{\flushright{IPPP/16/123\\}}
 \vspace{2cm}

\title{\boldmath On interpolations from SUSY to non-SUSY strings\\
and their properties\vspace{0.2cm}}

\author{ Benedict Aaronson\footnote{benedict.aaronson@durham.ac.uk}, Steven Abel\footnote{s.a.abel@durham.ac.uk}, Eirini Mavroudi\footnote{irene.mavroudi@durham.ac.uk}
\vspace{0.1cm}}
\affiliation{Institute for Particle Physics Phenomenology,
Durham University, South Road, Durham, DH1 3LE\vspace{0.5cm}}

\date{\today}

\begin{abstract}

\noindent The interpolation from supersymmetric to non-supersymmetric heterotic theories is studied, via the Scherk-Schwarz compactification of supersymmetric 
$6D$ theories to $4D$. A general modular-invariant Scherk-Schwarz deformation is deduced from the properties of the $6D$ theories at the endpoints, which 
significantly extends previously known examples. This wider class of non-supersymmetric $4D$ theories opens up new possibilities for model building. The full one-loop cosmological constant of such theories is studied as a function of compactification radius for a number of cases, and
the following interpolating configurations are found: two supersymmetric $6D$ theories related by a $T$-duality transformation, with intermediate $4D$ maximum or minimum at the string scale; 
a non-supersymmetric $6D$ theory interpolating to a supersymmetric $6D$ theory, with the $4D$ theory possibly having an AdS minimum; 
a ``metastable'' non-supersymmetric $6D$ theory interpolating via a $4D$ theory to a supersymmetric $6D$ theory. 
\end{abstract}

\maketitle

\def\beq{\begin{equation}}
\def\eeq{\end{equation}}
\def\beqn{\begin{eqnarray}}
\def\eeqn{\end{eqnarray}}
\def\half{{\textstyle{1\over 2}}}
\def\quarter{{\textstyle{1\over 4}}}

\def\calO{{\cal O}}
\def\calC{{\cal C}}
\def\calE{{\cal E}}
\def\calT{{\cal T}}
\def\calM{{\cal M}}
\def\calN{{\cal N}}
\def\calF{{\cal F}}
\def\calS{{\cal S}}
\def\calY{{\cal Y}}
\def\calV{{\cal V}}
\def\ibar{{\overline{\imath}}}
\def\chibar{{\overline{\chi}}}
\def\ttwo{{\vartheta_2}}
\def\tthree{{\vartheta_3}}
\def\tfour{{\vartheta_4}}
\def\ttwob{{\overline{\vartheta}_2}}
\def\tthreeb{{\overline{\vartheta}_3}}
\def\tfourb{{\overline{\vartheta}_4}}
\def\Str{{{\rm Str}\,}}

\def\bfell{{\boldsymbol \ell}}
\def\xx{\hspace{0.3cm}}
\def\xxl{\hspace{0.295cm}}
\def\xxh{\hspace{0.242cm}}
\def\yy{\hspace{0.115cm}}
\def\yyr{\hspace{-0.02cm}}

\def\qbar{{\overline{q}}}
\def\mm{{\tilde m}}
\def\nn{{\tilde n}}
\def\rep#1{{\bf {#1}}}
\def\ie{{\it i.e.}\/}
\def\eg{{\it e.g.}\/}

\newcommand{\newc}{\newcommand}
\newc{\gsim}{\lower.7ex\hbox{$\;\stackrel{\textstyle>}{\sim}\;$}}
\newc{\lsim}{\lower.7ex\hbox{$\;\stackrel{\textstyle<}{\sim}\;$}}

\newcommand{\red}[1]{\textcolor{red}{#1}}

\hyphenation{su-per-sym-met-ric non-su-per-sym-met-ric}
\hyphenation{space-time-super-sym-met-ric}
\hyphenation{mod-u-lar mod-u-lar--in-var-i-ant}


\def\inbar{\,\vrule height1.5ex width.4pt depth0pt}

\def\IC{\relax\hbox{$\inbar\kern-.3em{\rm C}$}}
\def\IQ{\relax\hbox{$\inbar\kern-.3em{\rm Q}$}}
\def\IR{\relax{\rm I\kern-.18em R}}
 \font\cmss=cmss10 \font\cmsss=cmss10 at 7pt
\def\IZ{\relax\ifmmode\mathchoice
 {\hbox{\cmss Z\kern-.4em Z}}{\hbox{\cmss Z\kern-.4em Z}}
 {\lower.9pt\hbox{\cmsss Z\kern-.4em Z}} {\lower1.2pt\hbox{\cmsss
 Z\kern-.4em Z}}\else{\cmss Z\kern-.4em Z}\fi}

\tableofcontents


\section{Motivation for studying interpolating models}
\label{sec:Motivation}

\noindent An important question in string phenomenology is how and when supersymmetry (SUSY) is broken. A great deal of effort has been devoted to  frameworks in which it is broken non-perturbatively in the supersymmetric effective field theory. Much less effort has been devoted to string theories that are non-supersymmetric by construction.

On the face of it, the trade off for the second option, is that non-supersymmetric string models do not have the stability properties of supersymmetric ones. However it can be argued that 
as long as the SUSY breaking is spontaneous and parametrically smaller than the string scale, the associated instability is under perturbative control \cite{Abel:2015oxa}. There is then  
no genuine moral, or even practical, advantage to the former more traditional option, since nature is not supersymmetric. Sooner or later, either route to the Standard Model (SM) will lead to runaway potentials for moduli that need to be stabilised. Indeed spontaneous breaking at the string level may even confer advantages in this respect, as discussed in ref.\cite{Abel:2016hgy}.  

Parametric control over SUSY breaking requires a generic method for passing from a non-superymmetric theory to a supersymmetric counterpart, under certain limiting conditions. The method that was studied in ref.\cite{Abel:2015oxa} is interpolation via compactification to lower dimensions, with SUSY broken by the Scherk-Schwarz mechanism \cite{scherkschwarz}. 
The two great advantages of interpolating models are that their compactification volumes can be tuned to make the cosmological constant arbitrarily small, and that some of them exhibit enhanced stability due to a one-loop cosmological constant that is exponentially suppressed with respect to the generic SUSY breaking scale \cite{Abel:2015oxa}. They can be viewed as natural and phenomenologically interesting extensions of the original observation in refs.\cite{Itoyama:1986ei,Itoyama:1987rc} that the $10D$ tachyon-free non-supersymmetric $SO(16)\times SO(16)$ model interpolates to the heterotic $E_8\times E_8$ model, via a Scherk-Schwarz compactification to $9D$. 

The general properties under interpolation of theories broken by the Scherk-Schwarz mechanism are not known. For example, what determines if the zero radius endpoint theory is supersymmetric? This paper focusses on the properties of  $4$-dimensional ($4D$) theories that interpolate between stable, supersymmetric $6D$ tachyon-free models. 
Three main results are presented.
\begin{itemize}
\item First, we derive and study the general form of the $6D$ endpoint theories, and show that their modular invariance properties derive directly from the Scherk-Schwarz deformation. This enables us to generalise the construction of modular invariant Scherk-Schwarz deformed theories by beginning with the $6D$ endpoint theory.

\item Second, we determine a simple criterion for whether a SUSY theory, broken by Scherk-Schwarz, will interpolate to a SUSY or a non-SUSY one at zero radius: the zero radius theory is non-supersymmetric, if and only if the Scherk-Schwarz acts on the gauge group as well as the space-time side.

\item Third, we undertake a preliminary survey (in the sense that the models we study only have orthogonal gauge groups) of some representative models that confirm these two properties, by examining their potentials and spectra.

\end{itemize}  

\begin{figure}
\begin{center}
\includegraphics[scale=0.75]{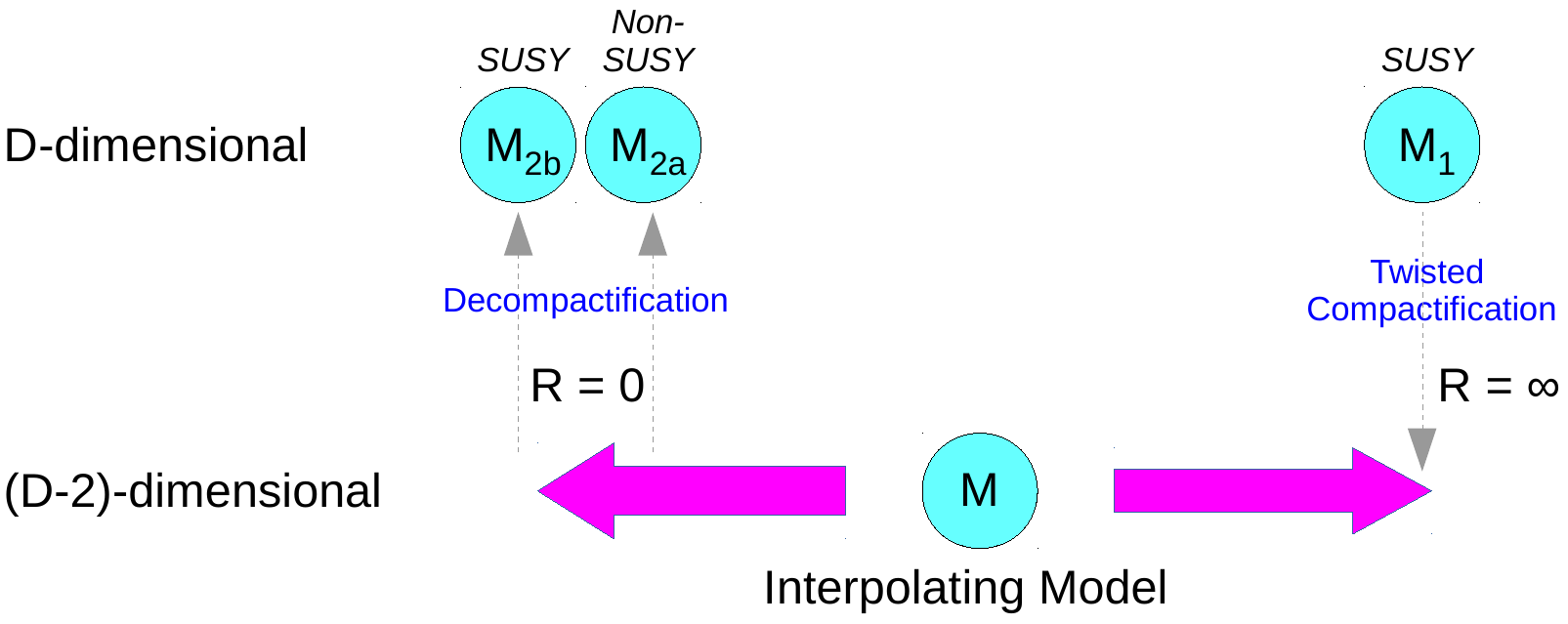}
\end{center}
\caption{\em The interpolation map between a $6D$ supersymmetric theory at infinite radius, $\mathcal{M}_1$ and its supersymmetric and non-supersymmetric, interpolated $6D$ duals, which are defined in the vanishing radius limit, with the vertical direction representing dimension. Whether or not SUSY is restored in the non-compact $r_1=r_2=r=0$, $6D$ model is determined by the structure of the Scherk-Schwarz action.}
\label{fig:Interpolation}
\end{figure}

The general framework for the interpolations are as shown in Figure \ref{fig:Interpolation}. Beginning with a supersymmetric $6D$  theory generically referred to as ${\mathcal M}_1$, 
the theory is compactified to a non-supersymmetric $4D$ theory ${\mathcal M}$ by adapting the Coordinate Dependent Compactification (CDC) technique first
 presented in refs.\cite{Kounnas:1989dk,Ferrara:1987es,Ferrara:1987qp,Ferrara:1988jx}.
This is the string version of the Scherk-Schwarz mechanism, which spontaneously breaks SUSY in the $4D$ theory with a 
gravitino mass of ${\cal O}(1/2r_i)$ where $r_i$ is the largest radius carrying a Scherk-Schwarz twist. (We will use ``CDC'' and ``Scherk-Schwarz'' interchangeably.) As usual it is the gravitino mass that is the order parameter for SUSY breaking:
it can be continuously dialled to zero at large radius where SUSY is restored and ${\mathcal M}_1$ regained. 

One of the main properties that will be addressed is the nature of the theory as the radius of compactification is taken to zero. This depends upon the precise details of the Scherk-Schwarz compactification, and indeed we will find that the presence or otherwise of SUSY at zero radius depend on the choice of basis vectors and structure constants defining the model. It is possible that the $4D$ theory interpolates to either a supersymmetric or a non-supersymmetric model (${\mathcal M}_{2a}$ or ${\mathcal M}_{2b}$ respectively).  Models of the latter kind 
correspond to a $6D$ theory in which SUSY is broken by discrete torsion \cite{Abel:2015oxa}. 

We begin in \S \ref{sec:CDC_Overview} by reviewing the basic formalism for interpolation. Section \ref{sec:CDC_Virasoro_Modified_Algebra} then presents the construction of $4D$ non-supersymmetric models as compactifications of $6D$ supersymmetric ones. The modification of the massless spectra in the decompactification and $r_i\rightarrow 0$ limits  (with the latter corresponding to the decompactification limit of a $6D$  $T$-dual theory) is analysed, in order to determine the nature of the theories at the small and large radii endpoints. Section  \ref{sec:Details of Cosmological Constant Calculation} discusses the technique for rendering the cosmological constant in an interpolating form, allowing it to be calculated across a regime of small and large radii. The modification of the projection conditions and massless spectrum by the choice of basis vectors and structure constants is made explicit, and based on these observations, in particular how the CDC correlates with modified GSO projections in the $6D$ endpoint theories, \S\ref{sec:most-general} then derives the general form of deformation within this framework, extending previous 
constructions. This more general formulation may prove to be useful for future model building. 

The conditions under which SUSY is preserved or broken at the endpoints of the interpolation are discussed in \S \ref{sec:SUSY Restoration and Energy Scales}.  
Particular focus is given to the constraints on the appearance of light gravitino winding modes in the zero radius limit. It is found that models in which the CDC acts only on the space-time side, are inevitably supersymmetric at zero radius, while models within which the CDC vector is non-trivial on the gauge side as well yield a non-supersymmetric model in the same limit. This analysis paves the way for a presentation in  \S \ref{sec:Results} of explicit interpolations (in terms of their cosmological constants) in particular models that display various different behaviours: namely we find examples of interpolation between two supersymmetric $6D$ theories via $4D$ theories with negative or positive cosmological constant; interpolation between a non-supersymmetric $6D$ theory and a supersymmetric one, with or without an intermediate $4D$ AdS minima; we also find examples of ``metastable'' non-supersymmetric $6D$ theories (by which we mean theories that have a positive cosmological constant with an energy barrier)  that can decay to supersymmetric ones.  

As mentioned, this paper follows on from a reasonably large body of work on non-supersymmetric strings that is nonetheless much smaller than the work on supersymmetric theories. 
Following on from the original studies of the ten-dimensional $SO(16)\times SO(16)$ heterotic string~\cite{SOsixteen},
there were further studies of the one-loop cosmological constants~\cite{Rohm,nonSUSYgauge,Itoyama:1986ei,Itoyama:1987rc,Moore,
 Dienes:1990ij,KutasovSeiberg,heretic,missusy,supertraces,Kachru:1998hd,KachSilvothers,
  Shiu:1998he,Iengo:1999sm,DhokerPhong,Faraggi:2009xy},
their finiteness properties~\cite{missusy,supertraces,Angelantonj:2010ic},
their relations to strong/weak coupling duality symmetries~\cite{Bergman:1997rf,julie1,julie2,Faraggi:2007tj}, and string landscape ideas~\cite{Dienes:2006ut,Dienes:2012dc}. 
The relationship to finite temperature strings was explored in refs.~\cite{finitetemp,AtickWitten,wasKounnasRostand,Kounnas:1989dk,earlystringpapersfiniteT}).
Further development of the Scherk-Schwarz mechanism in the string context was made
 in refs.\cite{Kiritsis:1997ca, Dudas:2000ff,
Scrucca:2001ni, Borunda:2002ra, Angelantonj:2006ut}.
Progress towards phenomenology within this class has been made in 
refs.~\cite{Lust:1986kj, Lerche:1986ae,Lerche:1986cx, nonSUSYgauge, Chamseddine:1988ck, Font:2002pq,Faraggi:2007tj,Blaszczyk:2014qoa,Angelantonj:2014dia,Blaszczyk:2015zta,Nibbelink:2016lzi}.
Related aspects concerning solutions to the large-volume ``decompactification problem'' were discussed in refs.\cite{Faraggi:2014eoa,Kounnas:2015yrc,Partouche:2016xqu,Kounnas:2016gmz,Abel:2016hgy}.
Non-supersymmetric string models have also been 
explored in a wide variety of other configurations~\cite{othernonsusy, Sagnotti:1995ga,Sagnotti:1996qj,Angelantonj:1998gj,Blumenhagen:1999ns,Sugimoto:1999tx,Aldazabal:1999tw,Angelantonj:1999xc,Forger:1999ev,Moriyama:2001ge,Angelantonj:2003hr,Angelantonj:2004yt,Dudas:2004vi,GatoRivera:2007yi}, including studies of the
relations between scales in various schemes~\cite{Antoniadis:1988jn, Antoniadis:1990ew, 
Antoniadis:1992fh,Antoniadis:1996hk,Benakli:1998pw,Bachas:1999es,Dudas:2000bn}. Some aspects of this study are particularly relevant to the recent work in refs.\cite{Florakis:2016aoi}. 
 
Note that here we will not elaborate on the properties of the non-supersymmetric $4D$ theory at radii of order the string length. As we will see, and as found in ref.\cite{Abel:2015oxa}, often there is a minimum in the cosmological constant at this point which suggests some kind of enhancement of symmetry at a special radius. (Indeed often it is possible to identify gauge boson winding modes that become massless at the minimum.) There is therefore the possibility of establishing connections to yet more non-supersymmetric $4D$ theories. Conversely one can ask if {\emph {every}} non-supersymmetric tachyon-free $4D$ theory can be interpolated to a supersymmetric higher dimensional theory. We comment on this and other prospects in the Conclusions in \S\ref{sec:Conclusions}.

\section{The cosmological constant and generalized Scherk-Schwarz construction}
\label{sec:Model_Construction}
\subsection{Overview}
\label{sec:CDC_Overview}
In this section, we revisit the calculation of the cosmological constant in the Scherk-Schwarzed theories, and in particular  derive a formulation for the partition function of interpolating models, that is useful for the later analysis. The discussion is a natural generalisation of the ``compactification-on-a-circle'' treatment of ref.\cite{Abel:2015oxa}, and as we shall see it ultimately leads to an improved and more general construction  for this class of theory.  

Let us begin by briefly summarising the implementation of the Scherk-Schwarz mechanism described in that work. 
As already mentioned, this is incorporated using a Coordinate Dependent Compactification (CDC) \cite{Kounnas:1989dk} of an initially supersymmetric $6D$  theory, namely the ${\cal M}_1$ model.  For our purposes it is useful to define it in the fermionic formulation, although any construction method would be applicable. In this formulation, the initial theory is defined by assigning boundary conditions to worldsheet fermions. If they are real, this is encapsulated in a set of 28-dimensional basis vectors, $V_i$, containing periodic or antiperiodic phases. The sectors of the theory are given by the set of $\overline{\alpha V}\equiv  {\alpha_i V_i} + \Delta $ where $\Delta \in \mathbb{Z} $ so that $\overline{\alpha V}\in [ -\frac{1}{2},\frac{1}{2}) $. We follow the usual the convention that $\alpha_i$ denotes the sum over spin structures on the $\alpha$ cycle. The spectrum of the theory at generic radius in any sector is determined by imposing the GSO projections governed by the vectors $V_i$ and a set of structure constants  $k_{ij}$, according to the KLST set of rules in 
refs.\cite{Kawai:1986ah,Kawai:1986va,Kawai:1986vd,Kawai:1987ew} (and equivalently ref.\cite{Antoniadis:1986rn}), which are summarised in Appendix, \ref{ffs}.

The model is then further compactified down to $4D$ on a ${\mathbb T}_2/{\mathbb Z}_2$ orbifold. In the absence of any 
CDC the result would simply be an ${\cal N}=1$ model resulting from an overall $(\mathbb{K}_3\times \mathbb{T}_2)/\mathbb{Z}_2$ compactification. The $\mathbb{K}_3$ in question corresponds to the $6D$ $\mathcal{N}=1$ theory in the fermionic construction in our examples. In theories of the type discussed in \cite{Abel:2015oxa}, in which the orbifold twist preserves SUSY, the twisted sectors have a supersymmetric spectrum, and therefore do not contribute to the cosmological constant, and thus the nature of the orbifold is unimportant. The CDC is implemented by introducing a deformation, described by 
 another vector $\mathbf{e}$, of shifts in the charge lattice that depend on the radii $r_{i=1,2}$ of the ${\mathbb T}_2$; this will be shown explicitly below.  
Under the CDC, the Virasoro generators of the theory are modified, yielding an extra effective projection condition, (in addition to the GSO projections associated with the $V_i$ from which the initial $\mathbb{K}_{3}$ is constructed), governed by ${\bf e}$, on the states constituting the massless spectrum of the $4D$ theory. The remaining massless states are then characterized by their charges under the $U(1)$ symmetry associated with ${\bf e}$. To qualify as a Scherk-Schwarz mechanism, this $U(1)$ symmetry has to include some component of the $R$-symmetry in order to distinguish bosons from fermions, thereby projecting out the gravitinos, and breaking spacetime SUSY. 

The effect of the CDC of course disappears in the strict $r \rightarrow \infty$ limit where the Kaluza-Klein (KK) spectrum becomes continuous, and the $6D$ endpoint model ${\cal M}_1$ is recovered. 
On the other hand as we shall see the CDC turns into another GSO projection vector in the $r_i \rightarrow 0$ limit, where states either remain massless or become infinitely massive. Upon  $T$-dualising, \begin{equation} r_i \rightarrow  \tilde{r}_i=1/r_i\, ,\end{equation} the $\tilde{r}_i \rightarrow \infty$ model becomes the non-compact theory ${\cal M}_{2}$, whose properties depend precisely on the form of ${\bf e}$.
 
The theories at the two endpoints can contain a different number of states and charges. Because ${\bf e}$ can overlap the gauge degrees of freedom, ${\cal M}_{2}$ will generically have a gauge symmetry that differs from that of ${\cal M}_1$, and possibly no SUSY. As we will see the two are in fact linked: if ${\cal M}_{2}$ is supersymmetric then the gauge group is the same as that of ${\cal M}_1$, if it is not, then the gauge group is different.

\subsection{CDC-Modified Virasoro Operators}
\label{sec:CDC_Virasoro_Modified_Algebra}
Let us now elaborate on the above description. The conventions for the fermionic construction are as in refs.\cite{Kawai:1986ah,Kawai:1986va,Kawai:1986vd,Kawai:1987ew} and for the CDC are as outlined in ref.\cite{Abel:2015oxa}, and summarised in Appendix \ref{ffs}. 
That is the unmodified Virasoro operators are defined as
\begin{equation}
\label{eqn:Virasoro_1}
    {{L_0}/\overline{L}_0}=\frac{1}{2}\alpha' p_{L/R}^2 + {\rm \text{oscillator contributions}}\, ,
\end{equation}
where, in terms of the winding and KK numbers, $n_i$ and $m_i$ respectively, the left- and right-moving momenta 
for a theory compactified on two circles of radii $r_{i = 1,2}$ take the unshifted form
\begin{equation}
\label{eqn:Momenta}
p_{L/R} \sim \left(\frac{m_i}{r_i} +\!/\!- n_ir_i \right)\, .
\end{equation}

Ultimately we wish to derive the largest possible class of deformations to the Virasoro operators that is compatible with modular invariance. This will turn out to be more general than those considered 
in refs.\cite{Ferrara:1988jx, Kounnas:1989dk}. In order to achieve this, we will now display the most general modification possible of the Virasoro operators under the Scherk-Schwarz action, along with a free parameter $m_{\mathbf{e}}$, which will ultimately be fixed by imposing modular invariance:
\begin{align}
\label{eqn:Virasoro_2}
{{L_0}'/\overline{L}'_0}&=~\frac{1}{2}\left[\mathbf{Q}_{L/R}-\mathbf{e}_{L/R}(n_1+n_2)\right]^2+\frac{1}{4}\left[\frac{m_1+m_e}{r_1} +\!/\!- n_1r_1\right]^2 \nonumber \\
&~~~~~~~~+\frac{1}{4}\left[\frac{m_2+m_e}{r_2} +\!/\!- n_2r_2\right]^2-1/\frac{1}{2}+ \mbox{ other oscillator contributions}\, ,
\end{align}
where the other oscillator contributions can be deduced from \ref{eqn: 2}, and where $\mathbf{Q}$ are the vectors of Cartan gauge and $R$-charges, defined by $\mathbf{Q} = {\bf N}_{\overline{\alpha V}} + \overline{\alpha V}$. As promised, the parameter $m_e$ will now be determined by modular invariance. The partition function of the modified theory is then expressed in terms of $q=e^{2 \pi i \tau}$ (where as usual the real and imaginary parts of $\tau$ are defined to be $\tau = \tau_1 + i \tau_2$):
\begin{equation}
\label{eqn:Partition_function}
{\cal Z}(\tau)~=~
   {\rm Tr}\, \sum_{m_{1,2}, n_{1,2}}   
   \mathtt{g}  \, \overline{q}^{\overline{L}'_0}q^{L'_0}\, .
\end{equation}
Modular invariance requires $L'_0 - \overline{L}'_0 \in \mathbb{Z}$. Given that the original supersymmetric theory is modular invariant (i.e. $L_0 - \overline{L}_0 \in \mathbb{Z}$) this can be used to determine a consistent $m_e$ as follows:
\begin{align}
\label{eqn:Virasoro_difference_2d_alphae}
L'_0 - \overline{L}'_0&=(m_1n_1 + m_2n_2) + \frac{1}{2}\left[\mathbf{Q}_{L}^2-\mathbf{Q}_{R}^2\right] + {(n_1+n_2)\, m_e}- \mathbf{e} \cdot \mathbf{Q}(n_1+n_2) +\mathbf{e}\cdot\mathbf{e}\frac{(n_1+n_2)^2}{2}\nonumber \\
&={{L_0}-\overline{L}_0}+{(n_1+n_2) m_e}- (n_1+n_2)\mathbf{e} \cdot \left[\mathbf{Q} - \mathbf{e}\frac{(n_1+n_2)}{2}\right]\, ,
\end{align}
where the dot products are Lorentzian.
Thus a KK shift of
\begin{equation}
\label{newGSO}
m_e=\mathbf{e\cdot Q}-\frac{1}{2}(n_1+n_2)\,\mathbf{e\cdot e}\, ,
\end{equation}
is {\it sufficient} to maintain modular invariance in the deformed theory. This matches the result of ref. \cite{Kounnas:1989dk}. The vector ${\mathbf e}$ then lifts the masses of states according to their charges under the linear combination $q_e={\bf e\cdot Q}$.
Restricting the discussion to half-integer mass-shifts imposes the constraint $\mathbf{e} \cdot\mathbf{e} = 1$ mod(2). 
Later on the partition function will be reorganised into sums over different values of $4m_e=0\ldots 3$ (as we restrict the study to $\frac{1}{2}$ phases in all examples, fractions of at most $\frac{1}{4}$ can arise in the GSO projections via odd numbers of overlapping $\frac{1}{2}$'s). So far these deformations are precisely those of
refs.\cite{Kounnas:1989dk,Ferrara:1987es,Ferrara:1987qp,Ferrara:1988jx}: once we consider the interpolation to the $6D$ theories, it will become clear how they can be made general. 

Note that level-matching is preserved by the CDC, but the mass spectrum is modified rather than the number of degrees of freedom contained within the theory, as required for a spontaneous breaking of  SUSY \cite{Kounnas:1989dk,Ferrara:1987es,Ferrara:1987qp,Ferrara:1988jx}.
It is clear from eq.(\ref{eqn:Virasoro_2}) that for zero winding modes ($n_i=0$), states for which $q_e=\mathbf{e \cdot Q} \neq 0\,\,\,{\rm mod}(1)$ become massive under the action of the CDC. Conversely all the zero winding states in the NS-NS sector remain unshifted by the CDC since they are chargeless. As described in ref.\cite{Abel:2015oxa} there may or may not be massless gravitinos depending on whether the effective projection $\mathbf{e \cdot Q} = 0\,\,\, {\rm mod}(1)$ is aligned with the other projections: this is in turn dependent on the choice of structure constant, so that ultimately the breaking of SUSY is associated with breaking by discrete torsion.  

\subsection{Details of Cosmological Constant Calculation}
\label{sec:Details of Cosmological Constant Calculation}
To evaluate the cosmological constant, at given radii $r_1 = r_2 = r$, one must integrate each $q^{M} \bar{q}^N$ term (weighted by its coefficient $c_{MN}$) in the total 1-loop partition function over the fundamental domain $\mathcal{F}$ of the modular group: \begin{equation}
\label{eqn:lambda}
\Lambda^{(D)} \equiv -\frac{1}{2} \mathcal{M}^{(D)}\int_{\mathcal{F}} \frac{d^2\tau}{{\tau_{2}}^{2}}\, \mathcal{Z}_{total}(\tau)\, ,
\end{equation}
where $D$ is the number of uncompactified spacetime dimensions (equal to 4 at all intermediate radii between the small and large radius $6D$ endpoint theories,  along which the cosmological constant will be evaluated), and ${\cal M}\equiv M_{\rm string}/(2\pi)=1/(2\pi \sqrt{\alpha'})$ is the reduced string scale. Henceforth ${\cal M}$ is set to 1; it can be reinserted by dimensional analysis at the end of the calculation if desired. 
The integral splits into  upper ($\tau_{2}>1$) and lower regions of the fundamental domain. Only terms for which $M=N$ can receive contributions from both regions, with the $\tau_{1}$ integral yielding zero in the upper region when $M \neq N$, enforcing level matching in the infra-red (but allowing contributions from unphysical proto-graviton modes in the ultra-violet as described in ref.\cite{Abel:2015oxa}).

At general radius the evaluation of the cosmological constant is complicated immensely by the fact that $M,N$ vary with $r_i$.
In order to make the evaluation tractable, the total partition function, $\mathcal{Z}_{total}(\tau)$, has to be rearranged into separate bosonic and fermionic factors as follows. It is convenient to define $n=(n_{1}+n_{2})$ and $\ell=(\ell_{1}+\ell_{2})$. Twisted sectors do not need to be considered in this implementation as, being supersymmetric, they do not contribute to the cosmological constant. In other words, the cosmological constant calculated without the orbifolding, is the same up to a factor of two, as the actual cosmological constant, as explained in detail in \cite{Kounnas:1989dk,Ferrara:1987es,Ferrara:1987qp,Ferrara:1988jx,Abel:2015oxa}. However, we will make further comments on twisted sectors later when we come to generalise the construction.  
We have\begin{equation}
\label{eqn: 38p}
{\cal Z}(\tau)=\frac{1}{\tau_{2} \eta^{22} \overline{\eta}^{10}}\sum_{\vec{\ell},\vec{n}}{\cal Z}_{ \vec{\ell},\vec{n}}\sum_{\alpha,\beta} \Omega_{{\ell},{n}}{\tiny\begin{bmatrix}
\alpha\\
\beta
\end{bmatrix}}\, ,
\end{equation}
where the Poisson-resummed partition function for the compactified complex boson is given by (see Appendix \ref{notation})
\begin{equation}
{\cal Z}_{\vec{\ell},\vec{n}} =~\frac{ r_1r_2}{\tau_2 \eta^2 \overline{\eta}^2}\sum_{{\vec{\ell},\vec{n}}}\exp \Big\{-\frac{\pi}{\tau_2}\left[r_1^2|\ell_1-n_1\tau|^2+r_2^2|\ell_2-n_2\tau|^2\right]\Big\}\, ,
\end{equation}
and the theta function products, each of which has characteristics defined by the sectors $\alpha, \beta$, with their respective CDC shifts, are
\begin{equation}
\label{eqn: 47}
\Omega_{{\ell},{n}}
{\tiny\begin{bmatrix}
\alpha \\
\beta \\
\end{bmatrix}}~=~ {\tilde{C}}^{\alpha,-n}_{\beta,-\ell}
\prod_{i_L }\vartheta{\tiny\begin{bmatrix}
\overline{\alpha\mathbf{V}_i-n\mathbf{e}_i}\\
-\beta\mathbf{V}_i +\ell\mathbf{e}_i\\
\end{bmatrix}}
\prod_{j_R}\overline{\vartheta}{\tiny\begin{bmatrix}
\overline{\alpha\mathbf{V}_j-n\mathbf{e}_j}\\
-\beta\mathbf{V}_j +\ell\mathbf{e}_j
\end{bmatrix}}\, ,
\end{equation}
where the conventions can be found in Appendix \ref{notation}.

In the above, the coefficients of the partition function are given by
\begin{equation}
\label{eqn: 45}
{\tilde{C}}^{\alpha,-n}_{\beta,-\ell}
~=~\exp\Big\{-2\pi i \left[n\mathbf{e}\cdot\beta{V}-\frac{1}{2}n\ell\mathbf{e}^2\right]\Big\}\, C^\alpha_\beta\, ,
\end{equation}
where $C^\alpha_\beta$ are the coefficients of the original theory before CDC, expressed in terms of the structure constants $k_{ij}$, and spin-statistic $s_i = V^{1}_i$, as in the original notation and Appendix \ref{ffs}, namely
\begin{equation}
\label{eqn: 32}
C_{\mathbf{\beta}}^{\mathbf{\alpha}}~=~\exp\left[2\pi i\left(\alpha s+\beta s+\beta_{i}k_{ij}\alpha_{j}\right)\right]\, .
\end{equation}

It is convenient to use the resummed version of this expression; certainly for the $q$-expansion this is the preferred method as it makes modular invariance explicit. This removes the $r_1 r_2$ prefactor and adds a factor of $\tau_2 $. The bosonic factor in the partition function $\mathcal{Z}_B(\tau)$ depend upon the radii of compactification, the winding and resummed KK numbers and the CDC induced shift in the KK levels, $m_e$, as follows:
\begin{multline}
{\cal Z}_{B_{ \vec{m},\vec{n},m_e}}=
 \frac{1}{\eta^{2} \bar{\eta}^{2}} \sum_{\vec{m},n_{1},k} q^{\frac{1}{4}\left(\frac{m_{1} + m_e}{r_{1}}+n_{1}r_{1}\right)^{2}+\frac{1}{4}\left(\frac{m_{2} + m_e}{r_{2}}+(n-n_{1}+4k)r_{2}\right)^{2}} \times \bar{q}^{\frac{1}{4}\left(\frac{m_{1} + m_e}{r_{1}}-n_{1}r_{1}\right)^{2}+\frac{1}{4}\left(\frac{m_{2} + m_e}{r_{2}}-(n-n_{1}+4k)r_{2}\right)^{2}} \, .
\end{multline}
The effective shift in the KK number, given by the requisite $m_e  \equiv {\bf e}\cdot ( {\bf Q} - n\frac{{\bf e}}{2} )$, 
arises from the choice of ${\tilde{C}}^{\alpha,-n}_{\beta,-\ell}
$, which gives an overall phase $e^{2\pi i \ell ({\bf e}\cdot ({\bf Q}-n{\bf e})- n {\bf e}^2/2)}$ in the partition function; as we shall see 
this shift in the KK number ultimately amounts to introducing a 
new vector $V_e\equiv {\bf e}$ in the non-compact $T$-dual theory at zero radius, combined with structure constants $k_{ei}=0$, $k_{ee}=1/2$. Note that this means in the 4D spectrum 
one may find states with 1/4-charges ${\bf e\cdot Q}=1/4,3/4$, that since they have $m_e\neq 0$, become infinitely massive in the zero radius limit.

In order to reorder the sum to do it efficiently, a projection in the ${\cal Z}_F$ on $\mathbf{Q}$ is now introduced to select possible values of $m_e$. Following the notation that $\beta_i$ represents the sum over spin structures, the parameter for this projection over the vector $\mathbf{e}$ will be called $\beta_e=0\ldots 3$. Thus overall, using the results in Appendix \ref{notation}, one can write,\begin{equation}
\label{mini1}
\mathcal{Z}_{total}(\tau) = \frac{1}{4}\frac{1}{\tau_{2} \eta^{22} \overline{\eta}^{10}}
\sum_{\stackrel{m_e=(0\ldots 3)/4}{\mbox{\tiny{$\vec{m},\vec{n}$}}}}{\cal Z}_{B_{ \vec{m},\vec{n},m_e}}
\sum_{\alpha,\beta,\beta_e} 
e^{2\pi i {\beta_e m_e} }
\Omega_{{n}}{\tiny\begin{bmatrix}
\alpha\\
\beta,\beta_e
\end{bmatrix}},
\end{equation}
where
\begin{equation}
\label{mini2}
\Omega_{{n}}
{\tiny\begin{bmatrix}
\alpha \\
\beta,\beta_e \\
\end{bmatrix}}~=~  {\tilde{C}}^{\alpha,-n}_{\beta,\beta_e} \prod_{i_L }\vartheta{\tiny\begin{bmatrix}
\overline{\alpha{V}_i-n\mathbf{e}_i}\\
{-\beta V_i - \beta_e\mathbf{e}_i}\\
\end{bmatrix}}
\prod_{j_R}\overline{\vartheta}{\tiny\begin{bmatrix}
\overline{\alpha{V}_j-n\mathbf{e}_j}\\
{-\beta{V}_j -\beta_e\mathbf{e}_j}
\end{bmatrix}}\, .
\end{equation}
Note that the phases in ${\tilde{C}}^{\alpha,-n}_{\beta,\beta_e}$ are precisely what is needed to cancel the contribution coming from the 
theta functions in $\Omega_n$, so that overall the  spectrum is merely shifted, with the GSO projections remaining independent of ${\bf e}$.

The bosonic contribution to the total partition function is independent of the fermionic sectors within the theory so  $\mathcal{Z}_B$ appears as a pre-factor to the sector sum for any given $m_e$. Conversely, the fermionic partition function is composed of terms that depend upon the boundary conditions of the fermions within the sectors $\alpha, \beta$ each of which is independent of the compactification radii. The advantage of this reordering is that one can therefore collect 16 representative factors, $n,\, 4 m_e={0..3\,\,\,  \mbox{mod(4)}}$,
\begin{equation} {\cal Z}_{F,n,m_e} =
\frac{1}{4}\sum_{\alpha\beta\beta_e}
e^{2\pi i {\beta_e m_e} }
\Omega_{{n}}{\tiny\begin{bmatrix}
\alpha\\
\beta,\beta_e 
\end{bmatrix}},\end{equation}
which are independent of the radii, and 16 respective ${\mathbb T}_2/{\mathbb Z}_2$ factors ($n,\, 4 m_e={0..3\,\,\,  \mbox{mod(4)}}$), which being independent of the internal degrees of freedom, depend only on the ${\mathbb T}_2$ compactification,
\begin{equation}
{\cal Z}_{B_{n,m_e}} =~\frac{1}{\eta^2 \overline{\eta}^2}\sum_{{\vec{m},n_1,k}}
q^{\frac{1}{4} \left( \frac{m_1 + m_e }{r_1} + n_1 r_1  \right)^2 + 
\frac{1}{4} \left( \frac{m_2 + m_e}{r_2} + (n-n_1+4k) r_2  \right)^2
}
 {\bar q}^{\frac{1}{4} \left( \frac{m_1 + m_e }{r_1} - n_1 r_1  \right)^2 
+\frac{1}{4} \left( \frac{m_2 + m_e }{r_2} - (n-n_1+4k) r_2  \right)^2
}\, .\end{equation}\\ 
The latter are radius dependent interpolating functions, analogous to the functions ${\cal E}_{0,1/2},{\cal O}_{0,1/2}$ in the simple circular case studied in ref.\cite{Abel:2015oxa}. We refer to the ${\cal Z}_{F,n,m_e}$ terms as `${\mathbb K}_3$ factors', since they involve only the \textit{internal} degrees of freedom of the $6D$ theory, and thus can be computed for all radii at the beginning of the calculation. The total partition function is then compiled by summing over the 16 $(n,m_e)$ sectors as 
\begin{equation}
\label{pf_reordered}
{\cal Z}(\tau)=\frac{1}{4}\frac{1}{\tau_{2} \eta^{22} \overline{\eta}^{10}}
\sum_{n,\,4m_e = 0..3}
{\cal Z}_{B,n,m_e}
{\cal Z}_{F,n,m_e}\, .
\end{equation}
To summarise, via the procedure of re-ordering the original sum \ref{eqn: 38p}, a projection on to different consistent $m_{e}$ values has been performed, such that a sum over $m_{e}$ can now be taken.

\subsection{The zero radius theory and a more general formulation of Scherk-Schwarz}

\label{sec:most-general}

An interesting aspect of the above approach is that in the small radius limit, that part of the spectrum with $m_e\neq 0$ mod(1) decouples and can be discarded, leaving the  partition function of the non-compact $6D$ theory at $r_i=0$.
Indeed, Poisson resumming on $n_1$ and $k$ gives 
\begin{equation} 
{\cal Z}_{B,n,m_e}\rightarrow  \sum_{{\vec{m}}}e^{-\left( \frac{(m_1+m_e)^2}{r_1^2} +\frac{ (m_2+m_e)^2}{r_2^2}\right)  \frac{\pi |\tau|^2}{\tau_2}} \frac{1}{4\tau_2 r_1 r_2}+\ldots \, ,
\end{equation}
where the ellipsis indicate terms that are further exponentially suppressed.
Thus the total untwisted partition function in the small radius limit can be expressed as
\begin{equation}
{\cal Z}(\tau)\rightarrow \frac{1}{16r_1r_2 }\frac{1}{\tau^2_{2} \eta^{22} \overline{\eta}^{10}}
\sum_{n}
{\cal Z}_{F,n,0}\, .
\end{equation}
Note that $1/(r_1r_2)$ is simply the expected volume factor of the partition function in the $T$-dual $6D$ theory. In conjunction with the fermionic component of the partition function, this then reproduces a $6D$ model with an additional basis vector ${\bf e}$, appearing in the sector definitions as $\alpha V-n {\bf e}$,   and with eq.(\ref{newGSO}) providing a 
new GSO projection, namely $ m_e={\bf{e}\cdot {\bf Q}}-n/2=0$ mod~(1). (The mod~(1) comes courtesy of the sum over $m_i$.) 

Upon inspection therefore, we are finding that eq.(\ref{newGSO}) {\it is actually} the GSO projection of an additional vector $V_e\equiv {\bf e}$ in  
the non-compact $6D$ theory. Beginning with the choice of ${\bf e}\cdot {\bf e}=1$,  one can infer that the $6D$ theory at zero radius for the examples we have been considering 
has structure constants $k_{ei}=0$ and $k_{ee}=1/2$, 
consistent with the modular invariance rules of KLST in refs.\cite{Kawai:1986ah,Kawai:1986va,Kawai:1986vd,Kawai:1987ew}. In fact identifying sectors as ${\alpha V}={\alpha_i V_i+\alpha_e V_e}$ with the sum over the spin structures on the $\mathbf{e}$ cycle as $\alpha_e=-n$ mod(2),  the entire partition function at zero radius is that of the 
$6D$ theory with the appropriate corresponding GSO phases,
\begin{equation}
\label{eq:be}
{\tilde{C}}^{\alpha,-n}_{\beta,\beta_e}
~=~ {C}^\alpha_\beta e^{2\pi i ( \beta_e k_{e j} \alpha_j -  \beta_i k_{ie} n  -  \beta_e k_{ee} n)}\, .
\end{equation}

Reversing the line of reasoning above, leads us finally to a generalisation of the construction of interpolating models based on the modular invariance 
of their endpoint $6D$ theories:
\begin{itemize}

\item First, define a $6D$ theory in terms of a set of vectors $V_i$, and {\it any} additional  $V_e\equiv {\bf e}$ vector that obeys the 
 $6D$ modular invariance rules of ref.\cite{Kawai:1986ah,Kawai:1986va,Kawai:1986vd,Kawai:1987ew},
together with a set of consistent structure constants $k_{ei}$ and $k_{ee}$.
(The $k_{ei}$ are then fixed by the modular invariance rules in the usual way.)

\item 
In theories that have an additional $\mathbb{Z}_2$ orbifold action ${\bf\hat{g}}$ on compactification to $4D$, 
$V_e\equiv {\bf e}$ is still constrained by the 
need to preserve mutually consistent  GSO projections, with the condition $\left\{ {\bf e\cdot Q} ,\, {\bf\hat{g}}\right\}=0$  (as in refs.\cite{Ferrara:1988jx, Kounnas:1989dk}
and discussed in ref.\cite{Abel:2015oxa}). 

\item  The partition function is then in the form of eqs.(\ref{mini1}),(\ref{mini2}) with coefficients as in eq.(\ref{eq:be}). The projection obtained by performing the $\beta_e$ sum determines the 
corresponding KK shift to be
\begin{equation}
\label{eq:genrule}
m_e={\bf e}\cdot {\bf Q} +(k_{ee}-{\bf e}^2) \,n - k_{ei} \alpha_i\, ,
\end{equation}
generalizing \ref{newGSO}.
\end{itemize} 

The last statment, namely that one may simply treat the Scherk-Schwarz action as another basis vector, leading to considerable generalisations, is one of the main results of the paper. In order to prove it, one may first Poisson-resum back to the original expression but retaining $\beta_e$, 
so that entire partition function is
\begin{equation}
\mathcal{Z}~=~
\frac{1}{4}\frac{1}{\tau_{2} \eta^{22} \overline{\eta}^{10}}
\sum_{\stackrel{m_e=(0\ldots 3)/4}{\beta_e}}
\sum_{\alpha,\beta,{\mbox{\tiny{$\vec{\ell},\vec{n}$}}}} 
\,
e^{2\pi i(\ell+\beta_{e})m_{e}}{\cal Z}_{\vec{\ell},\vec{n}}
\, {\tilde{C}}^{\alpha,-n}_{\beta,\beta_e} \prod_{i_L }\vartheta{\tiny\begin{bmatrix}
\overline{\alpha{V}_i-n\mathbf{e}_i}\\
{-\beta V_i - \beta_e\mathbf{e}_i}\\
\end{bmatrix}}
\prod_{j_R}\overline{\vartheta}{\tiny\begin{bmatrix}
\overline{\alpha{V}_j-n\mathbf{e}_j}\\
{-\beta{V}_j -\beta_e\mathbf{e}_j}
\end{bmatrix}}\, .
\label{eq:newform}
\end{equation}
Note that the sum over $m_e$ provides a projection that equates $\beta_e \equiv -\ell$ mod(1). Using the modular transformations for theta functions detailed in Appendix \ref{notation}, it is then straightforward to show that the partition function is
 invariant under 
$\tau \rightarrow \tau+1$ provided that   
\begin{equation}
e^{-i\pi\left(\mathbf{\overline{\alpha V}}-n\mathbf{e}\right)\cdot(2V_{0}+\mathbf{\overline{\alpha V}-n\mathbf{e}})}\tilde{C}{}_{\mathbf{\beta,\beta_{e}}}^{\mathbf{\alpha},-n}=\tilde{C}{}_{\mathbf{\beta-\alpha-}\delta_{i0}\mathbf{,}\beta_{e}+n}^{\mathbf{\alpha},-n},\label{eq:modi1}
\end{equation}
and invariant under 
$\tau \rightarrow -1/\tau$ provided that   
\begin{equation}
e^{-2\pi i\left(\mathbf{\overline{\alpha V}}-n\mathbf{e}\right)\cdot\left(\mathbf{\beta V}+\beta_{e}\mathbf{e}\right)}\tilde{C}{}_{\mathbf{\beta,\beta_{e}}}^{\mathbf{\alpha},-n}=
\tilde{C}{}_{-\mathbf{\alpha,}n}^{\beta,\beta_{e}}.\label{eq:modi2}
\end{equation}
This overall set of conditions is precisely that of KLST \cite{Kawai:1986ah,Kawai:1986va,Kawai:1986vd,Kawai:1987ew} with the original theory enlarged to include the vector $V_e\equiv {\bf e}$. $\square$
 
Note that these rules are significantly more general than those of refs.\cite{Kounnas:1989dk,Ferrara:1987es,Ferrara:1987qp,Ferrara:1988jx}, in which the choice  
\begin{equation}
\tilde{C}{}_{\mathbf{\beta,\beta_{e}}}^{\mathbf{\alpha},-n}=C_{\mathbf{\beta}}^{\mathbf{\alpha}}e^{-2\pi i\left(n\mathbf{e}\cdot\beta V+\beta_{e}n\frac{\mathbf{e\cdot e}}{2}\right)},
\end{equation}
corresponds to taking $k_{ei}=0$ and $k_{ee}=1/2$, in \ref{eq:be}. Now for example the CDC vectors are no longer restricted to obey ${\bf e}^2=1 $~mod(1), and 
moreover the KK shifts have additional sector dependence if $k_{ei}\neq 0$. We should add that, as well as being a generalisation, these rules simplify the construction of viable phenomenological models, because the  $\left\{ {\bf e\cdot Q} ,\, {\bf\hat{g}}\right\}=0$ condition can be implemented independently, with consistency then guaranteed with respect to 
all the other $V_i$ vectors \footnote{This is a somewhat subtle point because the basis in which the orbifold action is diagonal is not the same as the basis in which the Scherk-Schwarz action 
is diagonal. However the two act relatively independently on the partition function. This point is discussed in explicit detail in ref.\cite{ADMtocome}.}. 
One can also conclude that for consistency a theory that is Scherk-Schwarzed 
on an orbifold should contain additional sectors that are twisted under the action of both the orbifold and the Scherk-Schwarz -- i.e. twisted sectors that have non-zero $\alpha_e$. 
Of course $\alpha_e$ for such sectors has no association with any windings, but one finds that those sectors (which being twisted are supersymmetric) are required for consistency (anomaly cancellation for example).

\section{On SUSY Restoration}
\label{sec:SUSY Restoration and Energy Scales}
\subsection{Is the theory at small radius supersymmetric?}
\label{sec:Constructing_the_theory_at_small_radius}
Let us now move on to the conditions under which the endpoint theories exhibit SUSY. We will always consider models in which the theory at infinite radius is supersymmetric (as would be evidenced by the vanishing of the cosmological constant there) but we would like to determine whether or not SUSY is restored at zero  radius as well. In this section we develop arguments to address this question based on the existence or otherwise of massless gravitinos as $r_i \rightarrow 0$.

As usual the pure Neveu-Schwarz (NS-NS) sector, $\bf{0}$ gives rise to the gravity multiplet, $g_{\mu \nu}$ (the graviton), $\phi$ (the dilaton) and $B_{[\mu \nu]}$ (the two index antisymmetric tensor), from the states $\psi^{3,4}_{-\frac{1}{2}} \left|0\right\rangle_{R} \otimes X^{3,4}_{-1} \left|0\right\rangle_{L}$ in the notation 
of ref.\cite{Abel:2015oxa}. These states are chargeless under $\mathbf{e} \cdot \mathbf{Q}$ and no projection on them can occur, since the CDC vector is always zero in the $4D$ space-time dimensions $\psi^{3,4}$. Given the inevitable presence of the graviton, the SUSY properties of the theory are then dictated by the presence or absence of the R-NS gravitinos, namely \[ \Psi_\alpha^\mu  \equiv \left\{\psi^{3,4}_{0}\chi^{5,6}_{0}\chi^{7,8}_{0}\chi^{9,10}_{0} \right\}_\alpha  \left|0\right\rangle_{R} \otimes X^{34}_{-1} \left|0\right\rangle_{L}\, .\]  Their Scherk-Schwarz projections are determined purely by the Scherk-Schwarz  action on the right-moving degrees of freedom 

The spectrum is found from the expressions for the modified Virasoro operators in eq.(\ref{eqn:Virasoro_2}). For the non-winding gravitinos,  the shifted KK momentum becomes virtually continuous in the $r_i \rightarrow \infty$ limit and the full $6D$ gravitino state is inevitably recovered there. The scale at which SUSY is spontaneously broken by the CDC is set by the gravitino mass $1/2r_i$. As the compactification is turned on, the SUSY of the $6D$ theory is broken, and then towards the $r_i \rightarrow 0$ end of the interpolation, new gravitinos may or may not appear in the massless spectrum, perhaps heralding the restoration of SUSY at small radius as well.

To see if they do, consider how the CDC modifies the theories that sit at the endpoints of the interpolation. We denote by ${\bf Q}_{\psi}^0$ the charge of the lightest gravitino state at large radius. SUSY is exact even in the presence of ${\bf e}$, with the state ${\bf Q}_{\psi}^0$ being exactly massless, if both the first and second terms in the modified Virasoro operators of eq.(\ref{eqn:Virasoro_2}), namely
\begin{equation}
\label{eqn:Virasoro_term_1}
(\mathbf{Q}_{\psi}^0 - \mathbf{e}\,n)^2
\end{equation}
and
\begin{equation}
\label{eqn:Virasoro_term_2}
\left( \frac{m_i + \mathbf{e} \cdot \mathbf{Q}_{\psi}^0 -\frac{1}{2}n\mathbf{e^2}}{r_i} + n_i r_i \right)^2 \, ,
\end{equation}
vanish. (For convenience we continue for this discussion to use the original more restrictive rules of  refs.\cite{Kounnas:1989dk,Ferrara:1987es,Ferrara:1987qp,Ferrara:1988jx}; it would be trivial to 
extend the discussion to the more general rules of eq.(\ref{eq:genrule}).)
With $n_1 = n_2=0$, the first term receives no extra contribution due to the CDC. Furthermore, there is no winding contribution to the second term.  Therefore gravitinos that have  $\mathbf{e \cdot Q_{\psi}^0} = 0$ remain massless and indicate the presence of exact SUSY. Conversely, if the only remaining gravitinos have \begin{equation}\label{gravcon}
\mathbf{e \cdot Q_{\psi}^0} = \frac{1}{2}\, ,\end{equation}
their mass is $\frac{1}{2}\sqrt{\frac{1}{r_1^2} + \frac{1}{r_2^2}}$ and SUSY is spontaneously broken. 

Without loss of generality, one can consider SUSY breaking to amount to a conflict between $\mathbf{e}$ and a single basis vector, denoted by $V_{con}$. That is, $V_{con}$ constrains the gravitinos, while the remaining $V_i$ cannot project them out of the theory. In order for the above light (but not massless) gravitino to be the one that is left un-projected, the projections due to $\bf e$ and $V_{con}$ must disagree, that is the massive $\mathbf{e \cdot Q_{\psi}^0} = \frac{1}{2}$ state is retained by $V_{con}$ while the massless $\mathbf{e \cdot Q_{\psi}^0} = 0$ state is projected out. Again without loss of generality, it is always possible to choose $V_{con}$ so that the conditions are aligned; that is $V_{con}\cdot {\bf Q}_{\psi}^0=\frac{1}{2} \implies {\bf e}\cdot  {\bf Q}_{\psi}^0=\frac{1}{2}$. These modes are preserved (and have a mass $\sim \frac{1}{2r_i}$) while $V_{con}$ projects the massless $\mathbf{e \cdot Q_{\psi}} = 0$ modes out of the theory entirely.

Now consider the zero radius end of the interpolation, and denote the new would-be massless gravitino state by $\widetilde{\mathbf{Q}}_{\psi}$. Although a different state, it can be related to the infinite radius gravitino $\mathbf{Q}_{\psi}^{0}$ by a shift in the charge vector, induced by a potentially non-zero winding number; 
\begin{equation}
\label{eqn:Modified_Q_psi}
 \widetilde{\mathbf{Q}}_{\psi} = \mathbf{Q}_{\psi}^{0} - \mathbf{e}\,n\, .
\end{equation} 
As $r_i$ vanish, the spectrum associated with the winding modes becomes continuous, while the KK states become extremely heavy. As described in the previous section, the requirement that the KK term in eq.(\ref{eqn:Virasoro_term_2}) vanishes forms an effective projection that constrains the light states at zero radius, selecting the modes for which
\begin{equation}
\label{eqn:Modified_projection}
\mathbf{e}\cdot \mathbf{\widetilde{Q}}_{\psi}= \frac{n}{2}~~~ \text{mod(1),}
\end{equation}
where we will assume that $\mathbf{e \cdot e}=1$. 

It is clear from the relation between $\mathbf{\widetilde{Q}}_{\psi}$ and $\mathbf{Q}_{\psi}^0$ in eq.(\ref{eqn:Modified_Q_psi}) that the projection due to the CDC vector remains unchanged for any gravitino state there, since $\mathbf{e}^2 n \in \mathbb{Z}$; that is 
\begin{equation}
\mathbf{e} \cdot \mathbf{\widetilde{Q}}_{\psi} = \mathbf{e} \cdot \mathbf{Q}_{\psi}\, .
\end{equation}
This equation together with eqs.(\ref{eqn:Modified_projection}) and (\ref{gravcon}), imply that any gravitino of the spontaneously broken theory that becomes light at small radius must be an odd-winding mode.
Under the shift in $\mathbf{Q}$ given by eq.(\ref{eqn:Modified_Q_psi}), the $V_{con}$ projection constraining the gravitinos is
\begin{eqnarray}
\label{eqn:Modified_GSO_condition}
{\: V}_{con}\cdot \widetilde{\mathbf{Q}}_{\psi} &=&   V_{con} \cdot \mathbf{Q}^0_\psi -n\, V_{con} \cdot \mathbf{e}
 ~~\mbox{mod}\:(1)\, \nonumber \\&=&   \frac{1}{2} -n\, V_{con} \cdot \mathbf{e}
 ~~\mbox{mod}\:(1)\, .
\end{eqnarray}
For the effective projection in eq.($\ref{eqn:Modified_projection}$) to agree with the modified GSO condition in eq.(\ref{eqn:Modified_GSO_condition}) for  $n=odd$, we then require that \begin{equation}
\label{cond1}
V_{con} \cdot {\bf e}=0 \qquad {\mbox {mod~(1)}}. \end{equation}
  Eq.(\ref{cond1}) is a necessary condition for a model with SUSY spontaneously broken by the Scherk-Schwarz mechanism to have 
  massless gravitino states in both the infinite and zero radius limits. 
 
\subsubsection{SUSY restoration when the CDC vector has zero left-moving entries}
\noindent Let us see what it implies in a specific theory. Consider the basis vector set $\{V_0, V_1, V_2, V_4\}$, together with a CDC vector that is empty in its left-moving elements, the standard set up outlined in \cite{Abel:2015oxa}, in which the vectors $\{V_0, V_1, V_2\}$ project down to $6D$ SUSY with orthogonal gauge groups:
\begin{eqnarray}
\label{mod:output1}
V_0&=& - \mbox{${\scriptstyle \frac{1}{2}}$}[~11~111~111~ | ~1111~11111~111~111~11~111~]\nonumber\\ 
V_1&=& - \mbox{${\scriptstyle \frac{1}{2}}$}[~00~011~011~ | ~1111~11111~111~111~11~111~]\nonumber\\ 
V_2&=& - \mbox{${\scriptstyle \frac{1}{2}}$}[~00~101~101~ | ~0101~00000~011~111~11~111~]\nonumber\\ 
V_4&=& - \mbox{${\scriptstyle \frac{1}{2}}$}[~00~101~101~ | ~0101~00000~011~000~00~000~]\nonumber\\ 
{\bf e} &=&  - \mbox{${\scriptstyle \frac{1}{2}}$}[~00~101~101~ | ~0000~00000~000~000~00~000~]\, . 
\end{eqnarray}
A suitable and consistent set of structure constants is 
\begin{equation}
\label{eqn:vector_products_1}
k_{ij} =\left( \begin{array}{c}\mbox{$0$} \hspace{0.1cm}\mbox{$0$} \hspace{0.1cm}\mbox{$0$} \hspace{0.1cm}\mbox{$0$} \nonumber \\\mbox{$0$} \hspace{0.1cm}\mbox{$0$} \hspace{0.1cm}\mbox{$0$} \hspace{0.1cm}\mbox{$0$}  \nonumber \\\mbox{$0$} \hspace{0.1cm}\mbox{${\scriptstyle \frac{1}{2}}$}\hspace{0.1cm}\mbox{$0$} \hspace{0.1cm}\mbox{$0$} \nonumber \\\mbox{$0$} \hspace{0.1cm}\mbox{${\scriptstyle \frac{1}{2}}$}\hspace{0.1cm}\mbox{$0$} \hspace{0.1cm}\mbox{$0$} \nonumber \end{array}\right)\,. 
\end{equation}

Gravitinos are found in the $\overline{V_0 + V_1} = \frac{1}{2}[11~100~100~|~(0)^{20}]$ sector, with vacuum energies $[\epsilon_{R}, \epsilon_{L}] = [0, -1]$. The charge operator for the non-winding gravitinos in the initial (infinite radius) theory takes the same form as the sector vector itself. They have charges determined by $V_4$ that give the required ${\bf e\cdot Q}=1/2$ mod (1) for spontaneous SUSY breaking: the positive helicity states with this choice of structure constants  are 
\begin{equation}
\mathbf{Q}^0_{\psi} = \frac{1}{2}[1 \mbox{\,\,-\vspace{-0.1cm}1}\, {\mbox{\tiny$\pm$}}100~{\mbox{\tiny$\pm$}}100~|~(0)^{20}~]\, ,
\end{equation}
where the $\pm$ signs on the fermions are co-dependent.
It is clear from the vector overlap between $\mathbf{Q}^0_{\psi}$ and $V_4$ that the latter is playing the role of $V_{con}$ that constrains the gravitini states. (The structure constants have been chosen such that $V_2$ yields identical constraints.)
Whether or not any of the winding modes of the gravitinos are light at zero radius depends upon them satisfying the modified GSO projection condition of eq.(\ref{eqn:Modified_GSO_condition}):
\begin{equation}
\label{eqn:Modified_V_4_projection}
V_{4} \cdot \mathbf{\widetilde{Q}}_{\psi} = V_{4} \cdot \mathbf{Q}_{\psi}^{0} - V_{4} \cdot \mathbf{e} \,(n_1 + n_2)\, \;\;\;\; \text{mod (1)}\, .
\end{equation}
As we saw the two projections agree for the odd-winding modes of the $\widetilde{\mathbf{Q}}_{\psi}$ states since $V_{4} \cdot \mathbf{e} = 0 \,\, \text{mod(1)}$, 
and under the CDC, the charge vector for the small radius gravitino is 
\begin{equation}
\label{eqn:Modified_Q_psi_vector-0}
\widetilde{\mathbf{Q}}_{\psi} = \frac{1}{2}[1\mbox{\,\,-\vspace{-0.1cm}1} ~\,0\,0{\mbox{\tiny$\pm$}}1~0\,0{\mbox{\tiny$\pm$}}1~|~(0)^{20}~]\, .
\end{equation}
Note that non-zero right-moving charges of the small radius gravitino are on the $\omega^{3,4}$ and $\omega^{5,6}$ world-sheet degrees of freedom, and they no longer overlap the SUSY charges of the large radius theory. 

The appearance of gravitino states in the light spectrum in the zero radius limit of this theory reflects a general conclusion. If the left-moving elements of the CDC vector vanish, eq.(\ref{cond1}) is automatically satisfied. Any theory with a CDC vector acting purely on the space-time side becomes supersymmetric at zero radius since the projection always preserves the odd-winding modes of the gravitinos. The non-supersymmetric $4D$ theory at generic radius is therefore an interpolation between two supersymmetric theories quite generally in these cases, which sit at the zero and infinite radius endpoints. The supersymmetric nature of the zero radius theory will later be verified by the vanishing of the cosmological constant in the $r_i \rightarrow 0$ limit (Figure \ref{fig:plots_order_1_radii_increments_0.02_trivial_e_LM_Nb-Nf=28_r_only}), as presented in the following section. Note that the necessary cancellation between thousands of terms is highly non-trivial.

\subsubsection{Example of a CDC vector with non-zero left-moving entries}
\label{sec:CDC_vector_e_LM_non-zero}
\noindent Consider instead a theory composed of the same basis vector set as in eq.(\ref{mod:output1}), but now with a CDC vector containing non-zero left-moving entries: for example
\begin{eqnarray}
{\bf e} &=& \mbox{${\scriptstyle \frac{1}{2}}$}[~00~101~101~ | ~1011~00000~000~100~01~111~]\, .
\end{eqnarray}
Under the CDC, and for convenience of presentation dropping the $\pm$ signs, the charge vector for the odd-winding gravitino modes is modified to
\begin{equation}
\label{eqn:Modified_Q_psi_vector}
\widetilde{\mathbf{Q}}_{\psi} = \frac{1}{2}[1 1~001~001~|~1011~00000~000~100~01~111~]\, .
\end{equation}
As in the previous example the vector contains the same number of non-zero right-moving entries, but lying in different columns, so there is no contribution from eq.(\ref{eqn:Virasoro_term_1}) to the mass squared on the space-time side. However the non-zero left-moving elements now result in a non-zero contribution. Under the shift,
\begin{equation}
(\mathbf{Q}^0_{R},\mathbf{Q}^0_{L})^2 \rightarrow (\widetilde{\mathbf{Q}}_{R},\widetilde{\mathbf{Q}}_{L})^2 = (\mathbf{Q}^0_{R} + \mathbf{e}_{R},\mathbf{Q}^0_{L} + \mathbf{e}_{L})^2  \, ,
\end{equation}
any non-zero shift in $\mathbf{Q}^0_{L}$ will inevitably produce massive gravitinos since in the R-NS sector the charges of massless states must be zero mod (1) on the left-moving side.

We conclude that SUSY is restored at small as well as large radius if and only if the Scherk-Schwarz mechanism does not act on the gauge-side. Conversely if SUSY is broken at zero radius then so is the gauge symmetry.

\subsection{Formula for $N_b = N_f$?}
The nett bose-fermi number appears as the constant term in the parition function $\mathcal{Z}\supset (N_b - N_f)q^{0}\bar{q}^{0}+\ldots$. Thus,
the dominant terms in the one loop contribution to the cosmological constant are proportional to $(N_b - N_f)$ for the massless states \cite{Abel:2015oxa}, so non-supersymmetric models with an equal number of massless bosonic and fermionic states have an exponentially suppressed one-loop cosmological constant, and hence exhibit an increased degree of stability. Unfortunately it seems to be necessary to determine the full massless spectrum in order to deduce whether or not $N_b = N_f$. There appears to be no principle, or algebraically feasible generic procedure, for choosing the basis vectors $\{V_i\}$, the CDC vector $\mathbf{e}$, and the structure constants $k_{ij}$, that ensures that $N_b = N_f$. 

\section{Surveying the interpolation landscape}
\label{sec:Results}
We now turn to a survey of the different possible interpolations, in order to verify the rules derived in the previous sections, in particular those that govern the supersymmetry properties of the models. We should remark that in order to make the exercise computationally feasible, we will only use 1/2 phases so that the theories contain only large orthogonal gauge groups. As such, we are not here attempting to construct the SM, and the massless spectrum for each example will not be presented. (They can easily be determined using the rules in Appendix \ref{ffs}). Rather, studying the relationship between the cosmological constant and the radii of compactification exemplifies interpolation patterns between different types of model. Following the procedure outlined in Section \ref{sec:Details of Cosmological Constant Calculation}, the total partition function, $Z_{total}(\tau)$, truncated at an order $\mathcal{O}(q^2)$ in the $q$-expansion, which is computationally manageable while displaying the qualitative behaviour, is input in to the integral in \ref{eqn:lambda}, for a range of compactification radii between either ends of the interpolation range.

\subsection{Interpolation Between Two Supersymmetric Theories}
\subsubsection{Nb>Nf}
\begin{figure}
\begin{center}
\includegraphics[scale=0.8]{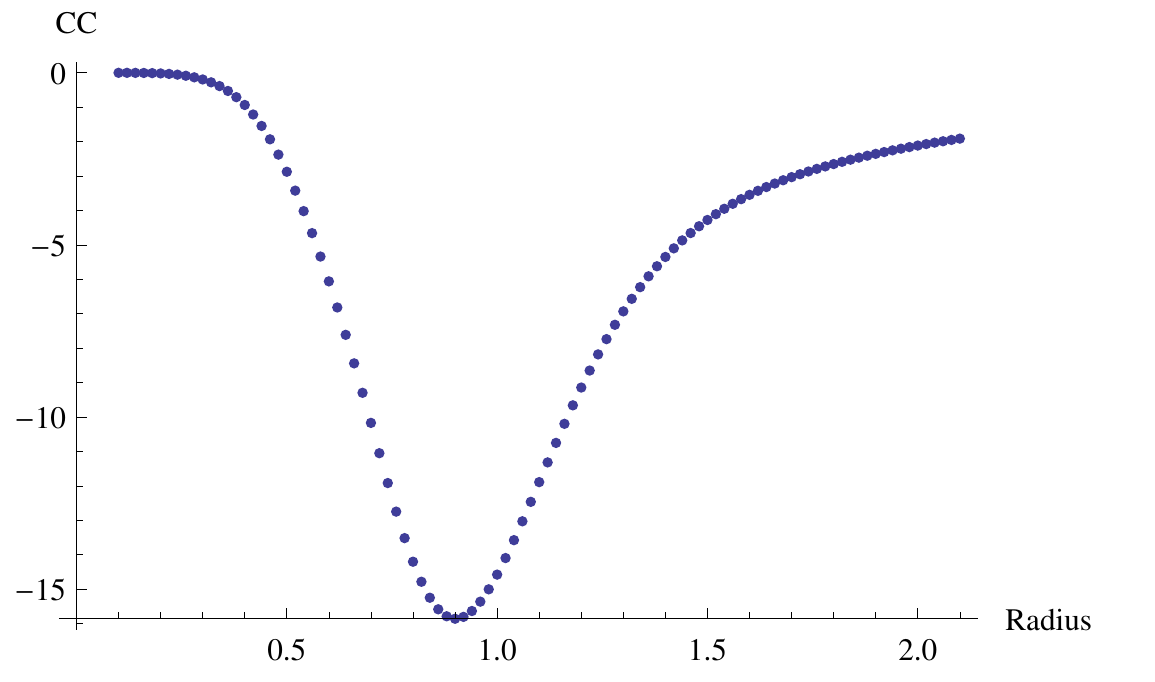}
\end{center}
\caption{\em Cosmological Constant vs. Radius, $r_{1}=r_{2}=r\in [0.1,2.1]$ with radius increments of $0.02$ for a model with $\mathbf{e}_{L} = $ trivial. $N_b-N_f=28$.}
\label{fig:plots_order_1_radii_increments_0.02_trivial_e_LM_Nb-Nf=28_r_only}
\end{figure}

\noindent Consider a theory containing $V_0$, $V_1$ and $V_2$ as in the above basis vector set in eqs.(\ref{mod:output1}), a modified $V_4$, an additional vector $V_5$, and a CDC vector that acts only on the space-time side:
\begin{eqnarray}
\label{mod:SUSY_to_SUSY_Nb>Nf}
V_4&=& - \mbox{${\scriptstyle \frac{1}{2}}$}[~00~000~000~ | ~0101~00000~000~011~00~000~]\nonumber\\ 
V_5&=& - \mbox{${\scriptstyle \frac{1}{2}}$}[~00~000~011~ | ~0101~11100~001~000~10~111~]\nonumber\\
{\bf e} &=& \mbox{${\scriptstyle \frac{1}{2}}$}[~00~101~101~ | ~0000~00000~000~000~00~000~] \, .
\end{eqnarray}
A suitable and consistent set of structure constants $k_{ij}$ is
\begin{equation}
\label{eqn:structure_constants_Nb>Nf_SUSY}
k_{ij} =\left( \begin{array}{c}\mbox{$0$} \hspace{0.1cm}\mbox{$0$} \hspace{0.1cm}\mbox{$0$} \hspace{0.1cm}\mbox{$0$} \hspace{0.1cm}\mbox{${\scriptstyle \frac{1}{2}}$} \nonumber \\\mbox{$0$} \hspace{0.1cm}\mbox{$0$} \hspace{0.1cm}\mbox{$0$} \hspace{0.1cm}\mbox{$0$} \hspace{0.1cm}\mbox{${\scriptstyle \frac{1}{2}}$}  \nonumber \\\mbox{$0$} \hspace{0.1cm}\mbox{${\scriptstyle \frac{1}{2}}$}\hspace{0.1cm}\mbox{$0$} \hspace{0.1cm}\mbox{${\scriptstyle \frac{1}{2}}$} \hspace{0.1cm}\mbox{${\scriptstyle \frac{1}{2}}$} \nonumber \\\mbox{$0$} \hspace{0.1cm}\mbox{$0$}\hspace{0.1cm}\mbox{${\scriptstyle \frac{1}{2}}$} \hspace{0.1cm}\mbox{${\scriptstyle \frac{1}{2}}$} \hspace{0.1cm}\mbox{${\scriptstyle \frac{1}{2}}$} \nonumber \\\mbox{${\scriptstyle \frac{1}{2}}$} \hspace{0.1cm}\mbox{${\scriptstyle \frac{1}{2}}$} \hspace{0.1cm}\mbox{$0$}\hspace{0.1cm}\mbox{$0$} \hspace{0.1cm}\mbox{${\scriptstyle \frac{1}{2}}$} \nonumber \end{array}\right)\, . 
\end{equation}
This model can be investigated using the general method presented in the previous section. $V_4$ plays the role of $V_{con}$, while $V_5$ respects its projections on the gravitinos. As discussed the interpolation is between two supersymmetric endpoints at both small and large radius. The cosmological constant takes a non-zero negative value with a minimum at intermediate values, and returns to zero at the two extremes, displayed in Figure \ref{fig:plots_order_1_radii_increments_0.02_trivial_e_LM_Nb-Nf=28_r_only}. 

\subsubsection{Nb<Nf}
\begin{figure}
\begin{center}
\includegraphics[scale=0.8]{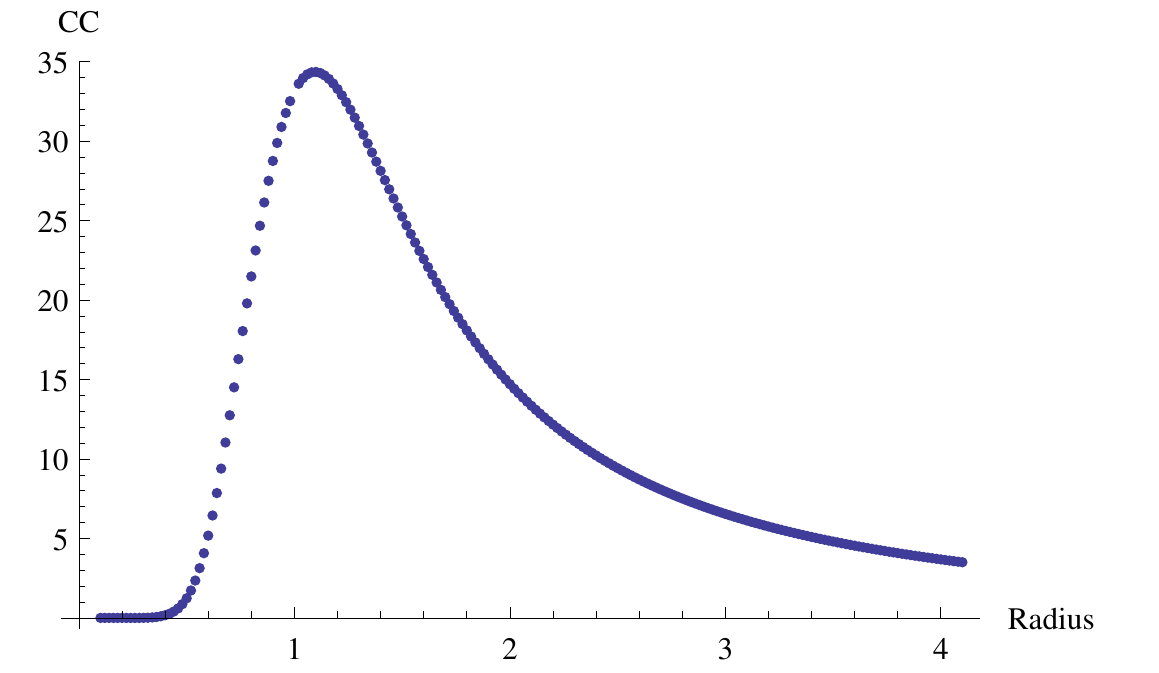}
\end{center}
\caption{\em Cosmological Constant vs. Radius, $r_{1}=r_{2}=r\in [0.1,4.1]$ with radius increments of $0.02$ for a model with $\mathbf{e}_{L} = $ trivial and $N_b-N_f=-228.$}
\label{fig:plots_order_1_radii_increments_0.02_trivial_e_LM_Nb-Nf=-228_r_only}
\end{figure}

\noindent A theory in which $N_b<N_f$ can be generated by a performing an alternative modification to the vectors $V_4, V_5$:
\begin{eqnarray}
\label{mod:SUSY_to_SUSY_Nb<Nf}
V_4&=& - \mbox{${\scriptstyle \frac{1}{2}}$}[~00~101~101~ | ~0101~00000~011~000~01~111~]\nonumber\\ 
V_5&=& - \mbox{${\scriptstyle \frac{1}{2}}$}[~00~000~011~ | ~0101~11100~010~110~00~011~]\nonumber\\
{\bf e} &=& \mbox{${\scriptstyle \frac{1}{2}}$}[~00~101~101~ | ~0000~00000~000~000~00~000~] \, ,
\end{eqnarray}
with the following structure constants:
\begin{equation}
\label{eqn:structure_constants_Nb<Nf_SUSY}
k_{ij} =\left( \begin{array}{c}\mbox{$0$} \hspace{0.1cm}\mbox{$0$} \hspace{0.1cm}\mbox{$0$} \hspace{0.1cm}\mbox{$0$} \hspace{0.1cm}\mbox{$0$} \nonumber \\\mbox{$0$} \hspace{0.1cm}\mbox{$0$} \hspace{0.1cm}\mbox{$0$} \hspace{0.1cm}\mbox{$0$} \hspace{0.1cm}\mbox{$0$}  \nonumber \\\mbox{$0$} \hspace{0.1cm}\mbox{${\scriptstyle \frac{1}{2}}$}\hspace{0.1cm}\mbox{$0$} \hspace{0.1cm}\mbox{$0$} \hspace{0.1cm}\mbox{${\scriptstyle \frac{1}{2}}$} \nonumber \\\mbox{$0$} \hspace{0.1cm}\mbox{${\scriptstyle \frac{1}{2}}$}\hspace{0.1cm}\mbox{$0$} \hspace{0.1cm}\mbox{${\scriptstyle \frac{1}{2}}$} \hspace{0.1cm}\mbox{$0$} \nonumber \\\mbox{$0$} \hspace{0.1cm}\mbox{$0$} \hspace{0.1cm}\mbox{$0$}\hspace{0.1cm}\mbox{$0$} \hspace{0.1cm}\mbox{$0$} \nonumber \end{array}\right)\, . 
\end{equation}
Similarly to the $N_b>N_f$ model with a exclusively non-trivial right-moving CDC vector, this model interpolates between two supersymmetric endpoints at both small and large radius, with the cosmological constant now taking a non-zero positive value at intermediate radii, displayed in Figure \ref{fig:plots_order_1_radii_increments_0.02_trivial_e_LM_Nb-Nf=-228_r_only}, corresponding to unstable runaway to decompactification at either end of the interpolation. 

\subsection{Interpolation from a Non-supersymmetric to a Supersymmetric Theory}
\subsubsection{Nb=Nf}
\begin{figure}
\begin{center}
\includegraphics[scale=0.8]{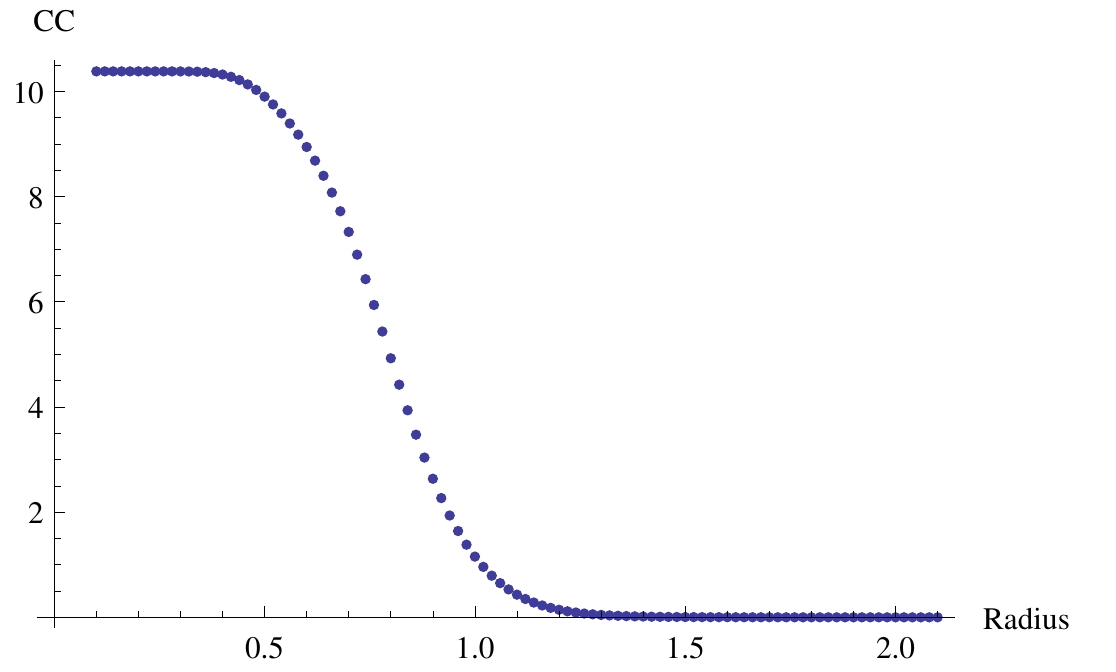}
\end{center}
\caption{\em Cosmological Constant vs. Radius, $r_{1}=r_{2}=r\in [0.1,2.1]$ with radius increments of $0.02$ for a model with $\mathbf{e}_{L} = $ non-trivial and $N_b = N_f$.}
\label{fig:plots_order_1_radii_increments_0.02_non_trivial_e_LM_Nb=Nf_r_only}
\end{figure}

\noindent A theory with Bose-Fermi degeneracy can be achieved with a theory comprised of the basis vector set in eq.(\ref{mod:output1}), plus a basis vector $V_5$ and CDC vector of the form
\begin{eqnarray}
V_5&=& - \mbox{${\scriptstyle \frac{1}{2}}$}[~00~000~011~ | ~0100~11100~000~111~10~011~] \, ,\nonumber \\
{\bf e} &=& \mbox{${\scriptstyle \frac{1}{2}}$}[~00~101~101~ | ~1011~00000~000~100~01~111~]\, ,
\end{eqnarray}
with $k_{ij}$ given by 
\begin{equation}
\label{eqn:structure_constants_2}
k_{ij} =\left( \begin{array}{c}\mbox{$0$} \hspace{0.1cm}\mbox{$0$} \hspace{0.1cm}\mbox{$0$} \hspace{0.1cm}\mbox{$0$} \hspace{0.1cm}\mbox{$0$} \nonumber \\\mbox{$0$} \hspace{0.1cm}\mbox{$0$} \hspace{0.1cm}\mbox{$0$} \hspace{0.1cm}\mbox{$0$} \hspace{0.1cm}\mbox{$0$}  \nonumber \\\mbox{$0$} \hspace{0.1cm}\mbox{${\scriptstyle \frac{1}{2}}$}\hspace{0.1cm}\mbox{$0$} \hspace{0.1cm}\mbox{$0$} \hspace{0.1cm}\mbox{$0$} \nonumber \\\mbox{$0$} \hspace{0.1cm}\mbox{${\scriptstyle \frac{1}{2}}$}\hspace{0.1cm}\mbox{$0$} \hspace{0.1cm}\mbox{$0$} \hspace{0.1cm}\mbox{$0$} \nonumber \\\mbox{$0$} \hspace{0.1cm}\mbox{$0$} \hspace{0.1cm}\mbox{${\scriptstyle \frac{1}{2}}$}\hspace{0.1cm}\mbox{$0$} \hspace{0.1cm}\mbox{$0$} \nonumber \end{array}\right)\, . 
\end{equation}
$N_b$ and $N_f$ are found to be equal despite the fact that the theory is non-supersymmetric (as can be seen by the absence of any massless gravitini in the spectrum). 
For models in which the CDC vector $\mathbf{e}$ is non-trivial in both the gauge and the global entries, the cosmological constant takes a non-zero value at small radius, while it vanishes exponentially quickly for large compactification scales, as displayed in Figure \ref{fig:plots_order_1_radii_increments_0.02_non_trivial_e_LM_Nb=Nf_r_only}.

\subsubsection{Nb>Nf}
\begin{figure}
\begin{center}
\includegraphics[scale=0.8]{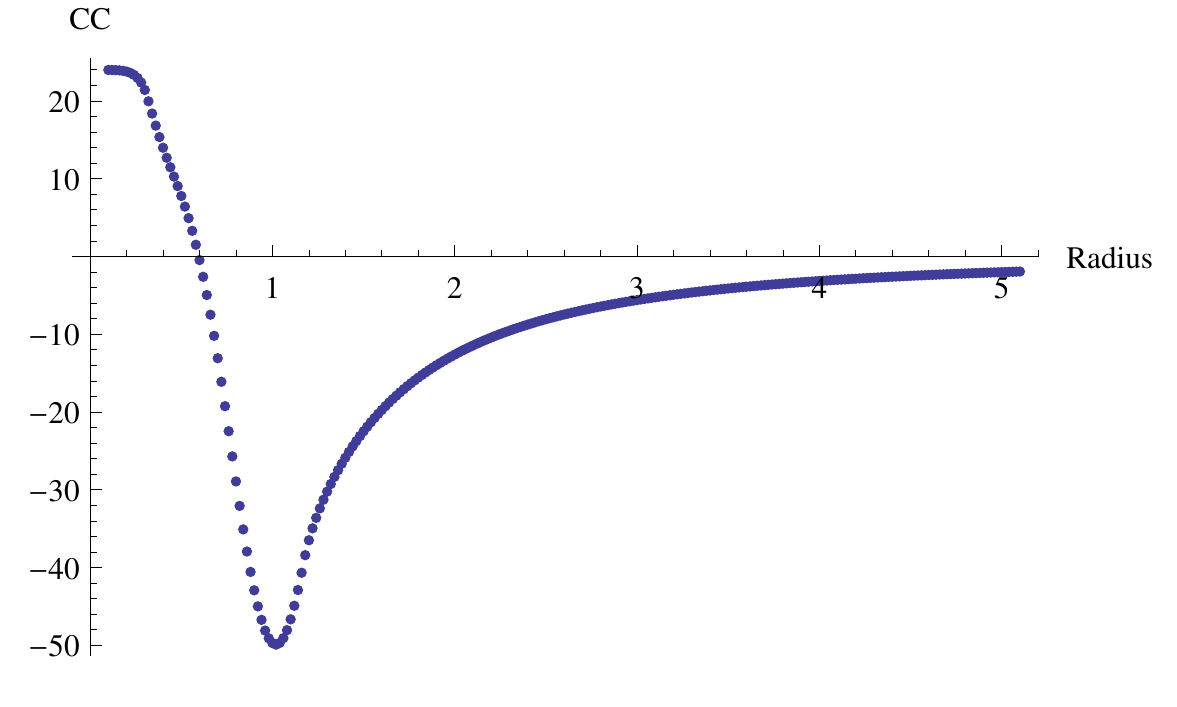}
\end{center}
\caption{\em Cosmological Constant vs. Radius, $r_{1}=r_{2}=r \in [0.1,5.1 ]$ with radius increments of $0.02$ for a model with $\mathbf{e}_{L} = $ non-trivial, and $N_b-N_f=192$.}
\label{fig:plots_order_1_radii_increments_0.02_non_trivial_e_LM_Nb-Nf=192_r_only}
\end{figure}

\noindent An interpolation from SUSY to non-SUSY in which $N_b>N_f$, can be achieved by taking the corresponding set of basis vectors in eqs.(\ref{mod:SUSY_to_SUSY_Nb>Nf}), but now with a CDC vector of the form
\begin{eqnarray}
{\bf e} &=& \mbox{${\scriptstyle \frac{1}{2}}$}[~00~101~101~ | ~0101~00000~000~110~11~011~]\, .
\end{eqnarray}
For models in which $N_b>N_f$, the cosmological constant reduces from a constant positive value at small radius reaching a negative minimum at approximately $r=1.0$ in string units. As the radius increases to $\infty$, the cosmological constant tends to zero from negative values, consistent with the restoration of SUSY in the endpoint model, as displayed in Figure \ref{fig:plots_order_1_radii_increments_0.02_non_trivial_e_LM_Nb-Nf=192_r_only}. In this particular example, the turnover appears to be at precisely 1 string unit, suggesting that a winding mode is becoming massless at this point, enhancing the gauge symmetry.

\subsubsection{Nb<Nf}
\begin{figure}
\begin{center}
\includegraphics[scale=0.8]{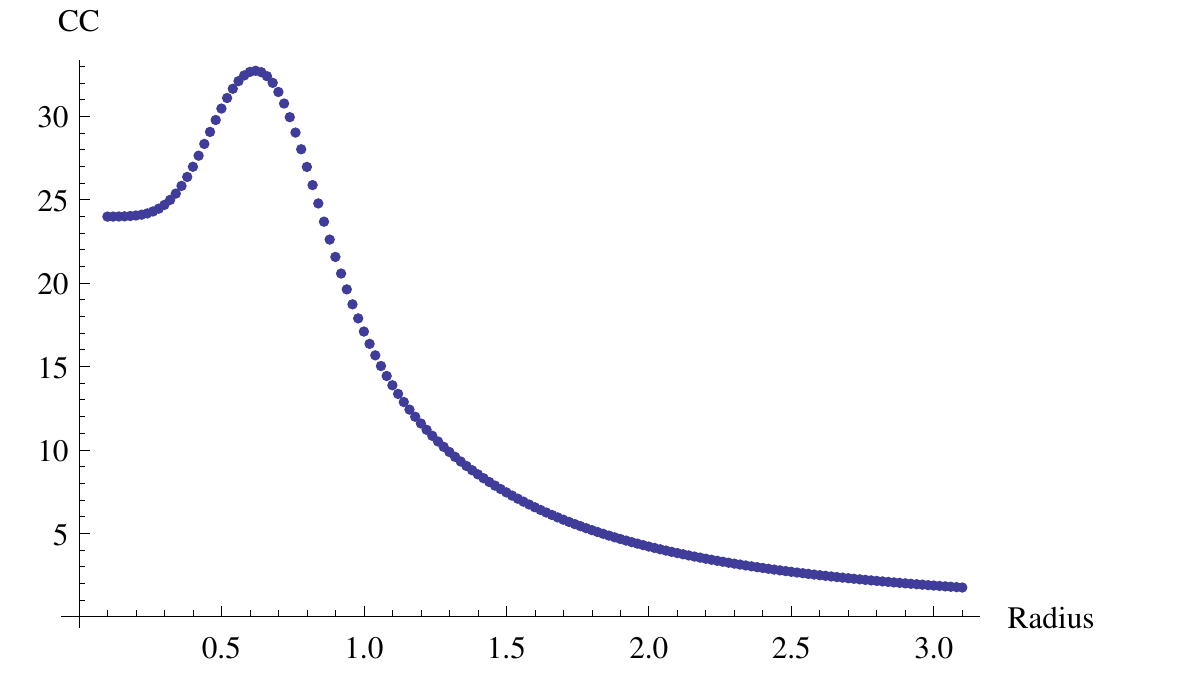}
\end{center}
\caption{\em Cosmological Constant vs. Radius, $r_{1}=r_{2}=r \in [0.1,3.1]$ with radius increments of $0.02$ for a model with $\mathbf{e}_{L} = $ non-trivial and $N_b-N_f=-64$.}
\label{fig:plots_order_1_radii_increments_0.02_non_trivial_e_LM_Nb-Nf=-64_r_only}
\end{figure}

\noindent Finally for a non-SUSY to SUSY interpolation with $N_b<N_f$, we take the model in eqs.(\ref{mod:SUSY_to_SUSY_Nb<Nf}) but now with a CDC vector of the form
\begin{eqnarray}
{\bf e} &=& \mbox{${\scriptstyle \frac{1}{2}}$}[~00~101~101~ | ~0101~00000~000~011~11~011~]\, .
\end{eqnarray}
The cosmological constant increases from a constant negative minimum at small radius, to a non-SUSY 6D theory at small radius and a SUSY 6D theory at infinite radius, as displayed in Figure \ref{fig:plots_order_1_radii_increments_0.02_non_trivial_e_LM_Nb-Nf=-64_r_only}.

\section{Conclusions}
\label{sec:Conclusions}

\noindent Following on from ref.\cite{Abel:2015oxa}, the nature of heterotic strings in the context of Scherk-Schwarz compactification has been investigated, 
with particular emphasis on their  
properties under interpolation. From the starting point of supersymmetric $6D$ theories in the infinite radius limit, Scherk-Schwarz compactification to $4D$ yields models 
that have $N_b \, \{ = \text{,} < \text{,} > \} \, N_f$, each possibility exhibiting different behaviours under interpolation. The behaviour of their cosmological constants was studied as a function 
of compactification radius, and it was found that theories can yield maxima or minima in the cosmological constant at intermediate values, as well as barriers with apparent metastability. 
The latter feature may have interesting phenomenological and/or cosmological applications.
The nature of the Scherk-Schwarz action, in particular whether or not it simultaneously acts to break the gauge group, dictates whether or not SUSY emerges in the $6D$ theory at zero radius. 

We studied the relation of the interpolating theory to the $6D$ theories that emerge at the end-points of the interpolation, and made the novel observation that the Scherk-Schwarz action descends from an additional GSO projection in the $6D$ zero radius endpoint theory. This allowed us to use the modular invariance constraints of the $6D$ theory to derive a more general class of Scherk-Schwarz compactification.

The aim in this work has been to establish the general features of interpolating models, relating higher, $D$-dimensional models to $(D-d)$-dimensional compactified models. It is conceivable that very many non-supersymmetric tachyon-free $4D$ models can be interpolated to higher dimensional supersymmetric ones. This would imply the existence of a formal relation between the process of interpolation, and the restoration of SUSY. Looking forward, it may not be possible to show that {\it every} non-supersymmetric theory is related to a supersymmetric counterpart via the process of interpolation. However, it seems possible that such a relation may always hold for the particular class of theories in which SUSY is broken by discrete torsion, as in ref.\cite{Blaszczyk:2014qoa} for example. 

A goal for future work would be to establish relationships, of the type found in this study, between additional lower dimensional, non-supersymmetric models, ideally of greater phenomenological appeal, and their supersymmetric counterparts. If it can be shown that non-supersymmetric models generically relate to supersymmetric theories in this way, interpolation could be used as a tool with which to relate many tachyon-free non-supersymmetric string theories to their supersymmetric siblings. Thus it would be possible to locate non-supersymmetric models within the larger network of string theories extending previous work in this direction. \\

\noindent {\bf Acknowledgements}: We are extremely grateful to Keith Dienes, Emilian Dudas and Herv\'e Partouche for many interesting discussions and comments. SAA would like to thank the \'Ecole Polytechnique for hospitality extended during this work. 

\newpage

\appendix

\section{~Notation and conventions for partition functions}
\label{notation}
\setcounter{footnote}{0}

The basic $\eta$ and $\vartheta$  functions are as given in \cite{Abel:2015oxa}. For convenience we will here reproduce the required generalizations of these functions. The more general theta functions with characteristics are defined as 
\begin{eqnarray}
\vartheta{\tiny\begin{bmatrix}
a\\
b
\end{bmatrix}}(z,\tau) & \equiv & \sum_{n=-\infty}^{\infty} e^{2\pi i(n+a)(z+b)} \, q^{(n+a)^{2}/2}\nonumber\\ 
 & = & 
e^{2\pi iab}\, 
    \xi^{a}\, 
    q^{a^{2}/2}\, \vartheta(z+a\tau+b,\tau)~;
\label{eq:jacobi}
\end{eqnarray}
of course these functions have a certain redundancy, 
depending on only $z+b$ rather than $z$ and $b$ separately.
In general, the functions in Eq.~(\ref{eq:jacobi}) have modular transformations
\beqn 
\vartheta{\tiny\begin{bmatrix} a\\ b \end{bmatrix}}(z,-1/\tau) &=& 
     \sqrt{-i\tau}\, e^{2\pi i a b}  e^{i \pi \tau z^2} 
 \vartheta{\tiny\begin{bmatrix} -b\\ a \end{bmatrix}}(-z\tau, \tau)~,  \nonumber\\
\vartheta{\tiny\begin{bmatrix} a\\ b \end{bmatrix}}(z,\tau+1) &=& 
e^{-i\pi (a^2 + a)}
\vartheta{\tiny\begin{bmatrix} a \\ a+b+1/2 \end{bmatrix}}(z,\tau)~.   
\label{modulartransfs}
\eeqn 
To evaluate the cosmological constant from the partition function in \S Section \ref{sec:Details of Cosmological Constant Calculation}, we require the following $q$-expansions:
\beqn
\label{eqn: 74}
\eta(\tau) & \sim & q^{1/24}+\ldots\nonumber \\
\vartheta{\tiny\begin{bmatrix} 0\\ 0 \end{bmatrix}}(0,\tau) & \sim & 1+2q^{1/2}+\ldots\nonumber \\
\vartheta{\tiny\begin{bmatrix} 0\\ 1/2 \end{bmatrix}}(0,\tau) & \sim & 1-2q^{1/2}+\ldots\nonumber \\
\vartheta{\tiny\begin{bmatrix} 1/2\\ 0 \end{bmatrix}}(0,\tau) & \sim & 2q^{1/8}+\ldots\nonumber \\
\vartheta{\tiny\begin{bmatrix} 1/2\\ 1/2 \end{bmatrix}}(0,\tau) & = & 0~.
\eeqn

Regarding partition functions, the expression for the compactified bosonic component of the partition function is given in \cite{Abel:2015oxa}. Here we will need the expression for the untilted torus in terms of radii $r_1$, $r_2$. 
The Poisson-resummed partition function is given by
\begin{equation}
    \mathcal{Z}_{\bf B}{\tiny\begin{bmatrix}
    0\\
    0 \end{bmatrix}}(\tau)~=~{\cal M}^2 \frac{r_1r_2}{{\tau_{2}}|\eta(\tau)|^{4}}\sum_{\mathbf{n},\mathbf{m}}\exp\left\{ -\frac{\pi}{\tau_{2}}r_1^2|m_{1}+n_{1}\tau|^{2}-\frac{\pi}{\tau_{2}}r_2^2|m_{2}+n_{2}\tau|^{2}\right\} .
    \end{equation} 
Each internal complex fermion degree of freedom contributes to the partition function depending on its world sheet boundary conditions,
$v\equiv \overline{\alpha V}_i $ and $u\equiv \beta V_i $, as
\begin{eqnarray}
\label{trace-formula}
\mathcal{Z}_{u}^{v} & = & \mbox{Tr}\left[q^{\hat{H}_{v}}e^{-2\pi iu\hat{N}_{v}}\right]\nonumber \\
& = & q^{\frac{1}{2}(v^{2}-\frac{1}{12})}\prod_{n=1}^{\infty}(1+e^{2\pi i(v\tau-u)}q^{n-\frac{1}{2}})(1+e^{-2\pi i(v\tau-u)}q^{n-\frac{1}{2}})\nonumber \\
& = & e^{2\pi iuv}\,
\vartheta{\tiny\begin{bmatrix}
v\\
-u
\end{bmatrix}}(0,\tau)/ {\eta(\tau)}~.
\end{eqnarray}

\section{~Conventions and spectrum of the fermionic string}
\label{ffs}

In this paper, the free-fermionic construction~\cite{Kawai:1986ah,Antoniadis:1986rn,Kawai:1987ew}
serves as the anchor underpinning our models.

In the free-fermionic construction, all world-sheet conformal anomalies are cancelled through the introduction of
free real world-sheet fermionic degrees of freedom.   
In the particular examples that we will be considering (which begin in $6D$), there are 8 right-moving and 20 left-moving complex Weyl fermions on the world-sheet. Models are defined by the phases acquired under parallel transport around 
non-contractible cycles of the one-loop world-sheet, 
\begin{align}
\label{eqn: 0.3}
\mathbf{1}: ~~~f_{i_{R/L}}&\rightarrow~ -e^{-2\pi i v_{i_{R/L}}}f_{i_{R/L}} \nonumber \\
\tau: ~~~f_{i_{R/L}}&\rightarrow~ -e^{-2\pi i u_{i_{R/L}}}f_{i_{R/L}}\, ,
\end{align}
where $i_R=1,\dots,8$ and $i_L=1,\dots,20$, which we collect in vectors written as
\begin{align}
\label{eqn: 0.2}
v\equiv\{v_R; v_L\}&\equiv~\{v_{i_R}; v_{i_L}\} \nonumber\\
u\equiv\{u_R; u_L\}&\equiv~\{u_{i_R}; u_{i_L}\}\, ,
\end{align}
where $v_{i_R}, v_{i_L},u_{i_R}, u_{i_L} \in [-\frac{1}{2},\frac{1}{2})$. 
The spin structure of the model is then given in terms of a set of  basis vectors ${V}_i$~\cite{Kawai:1987ew}. 
In order to define consistent modular invariant models, the basis vectors must obey
\beqn
\label{eqn: 1.1}
m_jk_{ij}&=&0 ~~~~~~~~~~~~~~~ {\rm mod}\,(1) \nonumber\\
k_{ij}+k_{ji}&=&{V}_i\cdot {V}_j ~~~~~~~~~ {\rm mod}\,(1) 
\nonumber\\
k_{ii}+k_{i0}+s_i &=& \frac{1}{2}{V}_i\cdot {V}_i 
~~~~~~~ {\rm mod}\,(1)~, 
\eeqn
where the $k_{ij}$ are otherwise arbitrary structure constants that completely specify the theory, 
where $m_i$ is the lowest common denominator amongst the components of ${V}_i$,
and where $s_i\equiv V_i^1$ is the spin-statistics associated with the vector $V_i$. 
The basis vectors span a finite additive group $G=\sum_k\alpha_k{V}_k$ where 
$\alpha_k\in \lbrace 0,..., m-1\rbrace$, each element of which describes the boundary conditions
associated with a different individual sector
of the theory.  Within each sector $\overline{\alpha V}$,  the physical states are those
which are level-matched and whose fermion-number operators ${N}_{\overline{\alpha V}}$ 
satisfy the generalized GSO projections
\begin{equation}
\label{eqn: 3b}
{\: V}_i\cdot {N}_{\overline{\alpha V}}~=~\sum_jk_{ij}\alpha_j+s_i-{\: V}_i\cdot\overline{\alpha {\: V}} \:\: \mbox{mod}\:(1) ~~~~~ {\rm for~all}~~i~.
\end{equation}
The world-sheet energies associated with such states are given by 
\beq
\label{eqn: 2}
M^2_{L,R} ~=~\sum_{{\ell}}\left\lbrace E_{\overline{\alpha V^{\ell}}} + \sum_{q=1}^\infty\left[(q-\overline{\alpha V^{\ell}})\overline{n}_q^{\ell}+(q+\overline{\alpha V^{\ell}}-1)n_q^{\ell}\right]\right\rbrace 
           -\frac{(D-2)}{24}+\sum_{i=2}^D\sum_{q=1}^\infty qM_q^i 
\eeq
where $\ell$ sums over left- or right world-sheet fermions, 
where $n_q, \overline{n}_q$ are the occupation numbers for complex fermions,
where $M_q$ are the occupation numbers for complex bosons, 
and where $E_{\overline{\alpha V^{\ell}}}$ is the vacuum-energy contribution 
of the $\ell^{\rm th}$ complex world-sheet fermion:
\begin{equation}
\label{eqn: 9}
E_{\overline{\alpha V^{\ell}}}~=~\frac{1}{2}\left[(\overline{\alpha V^{\ell}})^2-\frac{1}{12}\right]\, .
\end{equation}

Moreover, the vector of $U(1)$ charges for each complex world-sheet fermion is given by
\begin{equation}
\label{eqn: 5}
\mathbf{Q}~=~{N}_{\overline{\alpha V}}+{\overline{\alpha V}}
\end{equation}
where ${\overline{\alpha V}}$ is $0$ for an NS boundary condition and $-\frac{1}{2}$ for a Ramond.

\end{document}